\documentclass[prd,twocolumn,nofootinbib,showpacs,superscriptaddress]{revtex4-1}
\usepackage{amsfonts}

\expandafter\let\csname equation*\endcsname\relax
\expandafter\let\csname endequation*\endcsname\relax
\usepackage{amsmath}

\usepackage{mathalfa, amssymb, amsthm}
\usepackage{bm}
\usepackage{dcolumn}
\usepackage{graphicx}
\usepackage{graphics}
\usepackage{latexsym}
\usepackage{rotating}
\usepackage{hyperref}
\usepackage[all]{hypcap} 
\usepackage{xspace} % Sensible space treatment at end of simple macros
\usepackage[usenames,dvipsnames]{color}
\usepackage{mathrsfs}
\usepackage{subfigure}
\usepackage{slashed}
\usepackage[utf8]{inputenc}

\usepackage{mathalfa,amssymb, amsthm}
\usepackage{graphicx}
\usepackage{amsmath}
\usepackage{amssymb}
\usepackage{slashed}
\usepackage{graphicx}
\usepackage{setspace}
\usepackage{fullpage}
\usepackage{enumerate}
\usepackage{braket}
\usepackage[T1]{fontenc}
\usepackage{verbatim}

\theoremstyle{definition}

\allowdisplaybreaks

\usepackage{ulem}
\usepackage{enumitem}
\setenumerate[1]{label=\Roman*.}
\setenumerate[2]{label=\Alph*.}
\setenumerate[3]{label=\roman*.}
\setenumerate[4]{label=\alph*.}

\definecolor {darkgreen}{rgb}{0.2,0.7,0.2}

\usepackage{etoolbox}

\makeatletter
\def\@mkboth#1#2{}
\newlength\appendixwidth
\preto\appendix{\addtocontents{toc}{\protect\patchl@section}}
\newcommand{\patchl@section}{%
  \settowidth{\appendixwidth}{\textbf{Appendix }}%
  \addtolength{\appendixwidth}{1.5em}%
  \patchcmd{\l@section}{1.5em}{\appendixwidth}{}{\ddt}%
}
\makeatother

\newcommand\be{\begin{equation}}
\newcommand\ba{\begin{eqnarray}}
\newcommand\ee{\end{equation}}
\newcommand\ea{\end{eqnarray}}

\newcommand\bw{\begin{widetext}}
\newcommand\ew{\end{widetext}}

\newcommand{\nn}{\nonumber}

\newcommand{\om}{\bar{m}}
\newcommand{\ow}{\bar{\omega}}

\newcommand{\edth}{\textnormal{\dh}}

\usepackage{empheq}

\begin{document}
\title{Second Order Perturbations of Kerr Black Holes: \\ Formalism and Reconstruction of the First Order Metric}

\author{Nicholas Loutrel}
\affiliation{Department of Physics, Princeton University, Princeton, NJ, 08544, USA}
%\ead{nloutrel@princeton.edu}

\author{Justin L. Ripley}
\affiliation{Department of Physics, Princeton University, Princeton, NJ, 08544, USA}

\author{Elena Giorgi}
\affiliation{Department of Mathematics, Princeton University, Princeton, NJ, 08544, USA}
\affiliation{Princeton Gravity Initiative, Princeton University, Princeton, NJ 08544, USA}

\author{Frans Pretorius}
\affiliation{Department of Physics, Princeton University, Princeton, NJ, 08544, USA}
\affiliation{Princeton Gravity Initiative, Princeton University, Princeton, NJ 08544, USA}

\date{\today}

%%%%%%%%%%%%%%%%%%%%%%%%%%%%%%%%%%%%%%%%%%%%%%%%%
\begin{abstract} 
   Motivated by gravitational wave observations
of binary black hole mergers, we present a procedure to compute
the leading order nonlinear gravitational wave interactions
around a Kerr black hole.
%
% JLR: I commented out the below as it reads more like
% 'introduction' material
%
%Following a merger,
%there is strong evidence that the late-time linear
%emission
%is described by the quasi-normal modes of the remnant black hole.
%These are defined within the
%context of first order perturbation theory,
%governed by the Teukolsky equation.
%When comparing the results of linear theory
%to a realistic signal, whether simulated or detected,
%one has to wait a sufficiently long time for nonlinearities
%present during the merger phase to ``die down'',
%such that the signal becomes dominated by a sum of the
%exponentially damped sinusoids of the ringdown.
%For a detected event, this excludes the loudest part
%of the ringdown that could otherwise be used to
%interpret this phase of the merger.
%To remedy this situation requires understanding
%black hole perturbations beyond linear order,
%and in this work we take first steps toward
%characterizing the second order perturbations of Kerr black holes.
%It is known that such perturbations can be described by
%a Teukolsky-type equation, with a source term that
%depends on the square of first order perturbed quantities.
%Generically, this quadratic source term will induce
%mode-mixing at second order, with mode amplitudes
%and phases governed by those of the first order modes.
We describe the formalism used to derive the 
equations for second order perturbations.
We develop a procedure that allows
us to reconstruct the first order metric
perturbation solely from knowledge of the solution
to the first order Teukolsky equation,
without the need of Hertz potentials.
Finally, we illustrate this metric
reconstruction procedure in the asymptotic
limit for the first order quasi-normal modes of Kerr.
In a companion paper \cite{numerics_paper} we present
a numerical implementation of these ideas.
\end{abstract}

%\pacs{04.25.-g,04.25.Nx}

%04.30.Db Wave generation and sources
% 04.50.Kd Modified theories of gravity
% 04.25.-g Approximation methods; equations of motion
%04.25.Nx Post-Newtonian approximation; perturbation theory; related approximations
%97.60.Jd Neutron stars

\maketitle

%shut this off before submission
%\tableofcontents 

%%%%%%%%%%%%%%%%%%%%%%%%%%%%%%%%%%%
%\newpage
%%%%%%%%%%%%%%%%%%%%%%%%%%%%%%%%%%%%%%%%%%%%
\section{Introduction}
The coalescence of binary black holes generally proceeds through three phases:
the inspiral, merger, and ringdown.
In the inspiral phase, the orbital velocity is typically small
compared to the speed of light,
and one can solve the field equations of general relativity (GR)
using the perturbative post-Newtonian approximation~\cite{Blanchet:2013haa}.
In the merger phase, where the gravitational waves from the binary achieve
their maximum amplitude, the nonlinearities of GR cannot be neglected,
and one usually has to solve the field equations
numerically~\cite{Bishop:2016lgv}.
Finally, the ringdown phase constitutes the response of the final
black hole and is believed to be well described by the
quasi-normal modes computed
using black hole perturbation theory~\cite{Berti:2009kk}.

The ringdown phase of the coalescence not only provides
us with useful information regarding the remnant of binary mergers,
it also gives us a means of testing the conjectured uniqueness
of black holes in GR.
Several properties of black holes are related to uniqueness:
the no-hair theorems,
stating that the only {\em stationary}
black hole solutions in asymptotically
flat 4-dimensional spacetime with known matter fields are the
3-parameter (mass, spin angular momentum, and electric charge)
Kerr-Newman family~\cite{Israel:1967za, Israel:1967wq, Carter:1971zc, hawking-uniqueness,1975PhRvL..34..905R};
Penrose's weak cosmic censorship conjecture that when
gravitational collapse occurs the spacetime exterior to the
black hole horizon is complete;
and the final state conjecture~\cite{penrose1982},
a special case of which is the conjectured nonlinear
stability of the Kerr-Newman solutions,
whereby all dynamical perturbations (however large)
are absorbed by the black hole or radiated away,
leaving behind another member of the Kerr-Newman family.

The uniqueness properties of black holes offer
many avenues for testing the dynamical, strong-field regime of GR.
Regarding the ringdown,
the black hole spectroscopy proposal~\cite{Dreyer:2003bv,Berti:2005ys,Berti:2018vdi, Berti:2016lat}
exploits that the three parameters of the remnant
(or two in an astrophysical setting where charge is expected to be insignificant)
uniquely determine the frequencies and decay constants of the infinitely
many quasi-normal modes (QNMs) of the black hole;
hence, measurement of multiple modes do not provide novel
information about the black hole,
but instead are constraints to test uniqueness.
This just scratches the surface of what is theoretically possible:
for a ringdown produced by a binary black hole merger, the small
set of parameters of the progenitor binary not only
uniquely determines the remnant parameters
(and hence the QNM complex frequencies),
but also all the ``initial'' amplitudes and phases of all the QNM modes
(this forms the basis of the proposal to coherently stack
multiple detected events to enhance the ability to search
for subdominant modes~\cite{Yang:2017zxs}).
Moreover, all nonlinear effects, such as mode-coupling at second order,
are also uniquely governed by the progenitor parameters.
If the nonlinear phase of ringdown can be understood quantitatively,
this regime of a merger will also be accessible to uniqueness tests.

We should note however that if our only goal were to confirm GR
using black hole mergers, the residual test~\cite{LIGOScientific:2019fpa}
is adequate and does not require us to understand
or interpret phases of a merger; all one needs are full waveforms
computed with enough accuracy that subtraction of a
``best-fit'' waveform from the data leaves a residual signal consistent
with noise in the detectors.
Though if such a test were to fail,
it would be crucial to have a detailed knowledge of which part
of the waveform led to the residual,
and what novel physics or astrophysics that might point to
(whether exotic alternatives to black holes,
black holes with ``hair'',
or the usual GR black holes embedded in a circumbinary environment
sufficiently massive to measurably alter the uniqueness constraints
an isolated binary is subject to).

Each quasi-normal mode of the ringdown is identified by three integers,
two $(l,m)$ describing the angular dependence of the modes,
and one $(n)$ describing the overtone~\cite{Berti:2009kk}.
Generally, the late time behavior of the ringdown phase
is dominated by the leading order $(l,m,n) = (2,2,0)$
quadrupole mode,
but higher order modes become relevant under particular circumstances. 
Higher angular modes have comparable decay time to the
$(l,m,n) = (2,2,0)$ dominant mode,
but are more efficiently activated in systems with inherent asymmetries,
such as an unequal mass binary
(i.e. mass ratio $q \ne 1$)~\cite{Forteza:2020hbw}.
The first evidence for a non-quadrupole mode in the inspiral
phase came from the recent merger event
GW190412~\cite{LIGOScientific:2020stg},
however this was not loud enough for a corresponding QNM to be detected.

Overtones generally decay faster than the $n=0$ fundamental modes,
and thus can only be detected at higher signal-to-noise ratios (SNR),
or possibly, as with nonlinear effects,
if the analysis can be extended closer to the merger phase.
Intriguingly,\cite{Buonanno:2006ui,Giesler:2019uxc}
showed that for a merger of comparable mass non-spinning black holes,
as consistent with GW150914,
the waveform from peak amplitude onward can be well-fit with linear
modes if a sufficient number of overtones are included in the ringdown model.
There are caveats with this analysis, but if it turns out to be sound,
then there is already some evidence for observation of the first overtone
of the quadrupole mode with GW150914~\cite{Isi:2019aib}.
One of these caveats is, because of the rapid decay of the overtones,
with low SNR (or low accuracy in the model) rapidly decaying
nonlinear features could be fit by overtones
and be erroneously ascribed to them.
The study in ~\cite{Giesler:2019uxc} gave some evidence that this
was not occurring in their fits,
however back of the envelope estimates suggest second-order
mode coupling should be visible at comparable levels to the
higher overtones they included.
Without a detailed model of how the remnant black hole is
``excited'' during a merger to offer predictions for the various
components of the ringdown, rather than fitting,
it would be difficult to disentangle nonlinearity from overtones. 

Most analyses of the ringdown of black holes stop at
first order in perturbation theory. 
% FP: not sure the following statement fits in this paragraph
%However, such studies do not establish the validity of black hole perturbation theory outside of the linear regime. 
In generic perturbative problems,
second order perturbations are sourced by the square of
first order perturbations, constituting the leading order nonlinear effects.
This holds true for black hole perturbation theory. 
Historically, second order black hole perturbation theory
was originally considered~\cite{Gleiser:1995gx, Gleiser:1996yc}
to extend the close-limit approximation to
black hole mergers~\cite{Price:1994pm}.
These second order calculations were later applied
in the context of quasi-normal modes of
Schwarzschild black holes, where it was found that the
second order amplitudes could be as much as ten percent
of the first order
amplitudes~\cite{Nakano:2007cj, Ioka:2007ak, Pazos:2010xf}.
A rigorous proof of the stability of fully nonlinear perturbations
of a Schwarzschild black hole is only known restricted to a symmetry class
\cite{Klainerman:2017nrb}.
More recently, second order perturbation theory has been employed in
the self-force formalism as a necessity for computing accurate waveforms
for extreme mass ratio inspirals (EMRIs)
(see e.g.~\cite{Lousto:2008vw, Keidl:2010pm, Shah:2010bi, Gralla:2012db, vandeMeent:2017zgy}). 

This being said, much about second order perturbations of \emph{spinning}
black holes in the contexts of black hole ringdown and EMRI
remain open problems. 
A promising approach to study such perturbations was initiated
by Campanelli \& Lousto ~\cite{Campanelli:1998jv},
who employed the Newman-Penrose (NP)
formalism~\cite{Newman_Penrose_paper, Chandrasekhar_bh_book} to
derive an equation for second order gravitational wave
perturbations of Kerr black holes. 

In the NP formalism,
linear gravitational waves 
are described by the linear part of the Weyl scalar
$\Psi_{4}^{(1)}$.
(Here and below
we use the notation $f^{(n)}$ to denote the $n^{th}$-order perturbation
of $f$ about its background value $f^{(0)}$).
Campanelli \& Lousto's equation takes the form of a
Teukolsky equation for the second order $\Psi_{4}^{(2)}$ 
with a source term quadratic in first order perturbations.
The chief challenge to computing this source term in a practical manner
is that it depends on many more first order
geometric quantities than simply $\Psi_{4}^{(1)}$, and finding
the set consistent with the given $\Psi_{4}^{(1)}$ is what we refer to 
as {\em reconstruction}. (All the above can equivalently be
performed in terms of the NP scalar $\Psi_{0}$ instead
of $\Psi_{4}$).

An early method developed for reconstruction was given by
Chrzanowski~\cite{PhysRevD.11.2042, PhysRevD.11.2042}
(see also \cite{doi:10.1098/rspa.1979.0101}, and
~\cite{Whiting:2005hr} for a more recent review),
who showed that there exist ``Hertz'' potentials for gravitational
(and electromagnetic) perturbations in the Kerr background.
The gravitational Hertz potential solves the spin-weight $-2$
Teukolsky equation
(which we simply call the ``Teukolsky equation'' for brevity). 
Effectively then from a solution $\Psi$ to the Teukolsky equation
one can generate a perturbed metric that solves the linearized
Einstein equations about a Kerr background.
The complication with this approach is that while the Hertz potential
$\Psi$ solves the Teukolsky equation, it does not relate in a simple
way to the linearly perturbed Weyl scalar $\Psi_4^{(1)}$
(or $\Psi_0^{(1)}$).
Therefore, it is not possible to {\em directly} apply Chrzanowski's method if 
one wants to find the perturbed metric associated with 
a particular $\Psi_{4}^{(1)}$. 

A further drawback 
of Chrzanowski's method is that one is required to work 
in one of two radiation gauges, 
first described by Chrzanowski~\cite{PhysRevD.11.2042}
and later expanded on in~\cite{Price:2006ke}. These gauge conditions
can only be applied in Type II or more special spacetimes, 
and force particular conditions on the matter stress energy tensor.
This limits the Hertz potential method from directly
dealing with matter sources that do not satisfy those conditions, such as with EMRIs for example. Further, this 
technique cannot be applied at second order in perturbation theory to recover the second order metric 
perturbation, since the source terms coming from the first order perturbation act as 
effective matter sources that are not consistent with the conditions required for the radiation
gauges. 

Recently, a new approach was proposed in~\cite{Green:2019nam} 
to extend the Hertz potential approach to allow for arbitrary matter sources. The approach starts by 
giving an ansatz for the metric perturbation of the 
form $h_{ab} \sim \text{Re}[S^{\dagger} \Phi]_{ab} + x_{ab}$, 
where $S^{\dagger}$ is a second order differential operator, $\Phi$ is the Hertz potential,
and $x_{ab}$ is a ``correction'' tensor.
The first term on the right hand side is essentially Chrzanowski's method that
will give a linearized solution to the Einstein equations if the radiation
gauge conditions can be met; if not, $x_{ab}$ provides a correction 
proportional to the matter terms so that the net $h_{ab}$ 
does solve the linearized Einstein equations. 
Thus an additional
benefit of this procedure is that it allows for a path to calculating metric perturbations 
of the Kerr spacetime beyond linear order.

There are other workarounds to the above mentioned problems
(see e.g.~\cite{Lousto:2002em,Ori:2002uv,Merlin:2016boc}),
though there are also
procedures~\cite{Chandrasekhar_bh_book, Andersson:2019dwi} to directly
reconstruct the metric from $\Psi_4^{(1)}$, which avoid the use of
intermediate Hertz potentials.
In this work we describe a formalism building on the latter methods,
to compute the second order gravitational wave perturbation of
an arbitrary Type D spacetime that satisfies the vacuum Einstein equations.
The initial step is to write all first order NP quantities
(spin coefficients and Weyl scalars)
in terms of the background metric, and null tetrad projections of the first order 
metric perturbation and its gradients.
We use outgoing radiation gauge,
though note that in principle our method does not require such a gauge;
rather, it reduces the number of equations we need to solve in the end.

We then show how in this gauge, all first order NP quantities can be
derived from the solution of the
Teukolsky equation for $\Psi_4^{(1)}$, several additional null transport equations,
and some algebraic relations between spin coefficients
and the first order metric perturbation.
This then allows us to compute the source term necessary to
solve the Teukolsky equation for the second order
gravitational wave perturbation represented by $\Psi_4^{(2)}$.

At future null infinity in outgoing radiation gauge
$\Psi_4^{(2)}$ relates to the two polarizations of
the second order metric perturbation
($h^{(2)}_{\times}$ and $h^{(2)}_+$)
in exactly the same way $\Psi_4^{(1)}$ relates
to the linear metric \cite{Campanelli:1998jv}: 
\begin{align}
   \Psi_4^{(1,2)}
   =
-  \frac{1}{2}\left(
      \partial_t^2h_{+}^{(1,2)}
   -  i\partial_t^2h_{\times}^{(1,2)}
   \right)
   ,
\end{align}
Thus by reading off $\Psi_4^{(1)}$ and $\Psi_4^{(2)}$ at future null
infinity in outgoing radiation gauge we have a direct measure
of the relative magnitude of second order effects for a given choice of initial data.

%We note that our present procedure cannot be used to reconstruct the metric at second or higher order, which would be necessary if we wanted to compute the gravitational wave response at third (or higher) order. The reason is that outgoing radiation gauge can only be imposed in vacuum, or for matter (whether actual matter or effective matter coming from the first-order source terms) whose stress-energy tensor satisfies rather restrictive conditions (see Eq.(\ref{eq:matter-org}) below). Thus while our present metric reconstruction method avoids the use (and complications) of methods based on a Hertz-potential, it also has a more limited regime of applicability than, e.g. the recently proposed metric reconstruction procedure laid out by Green et. al.~\cite{Green:2019nam} \footnote{This being said, conceptually the direct metric reconstruction approach we adopt here does not require the use of outgoing radiation gauge, see for example Chandrasekhar \cite{Chandrasekhar_bh_book} and Andersson et. al. \cite{Andersson:2019dwi}. Outgoing radiation gauge is convenient due to its simplicity (essentially five metric components are set to zero, one more than would be expected for a linearized gauge in a generic spacetime background). With a less restrictive gauge, it should be possible to incorporate linearized matter sources with the direct metric reconstruction approach, although this is beyond the scope of the present paper.}.

\begin{figure}
\includegraphics[width=\columnwidth]{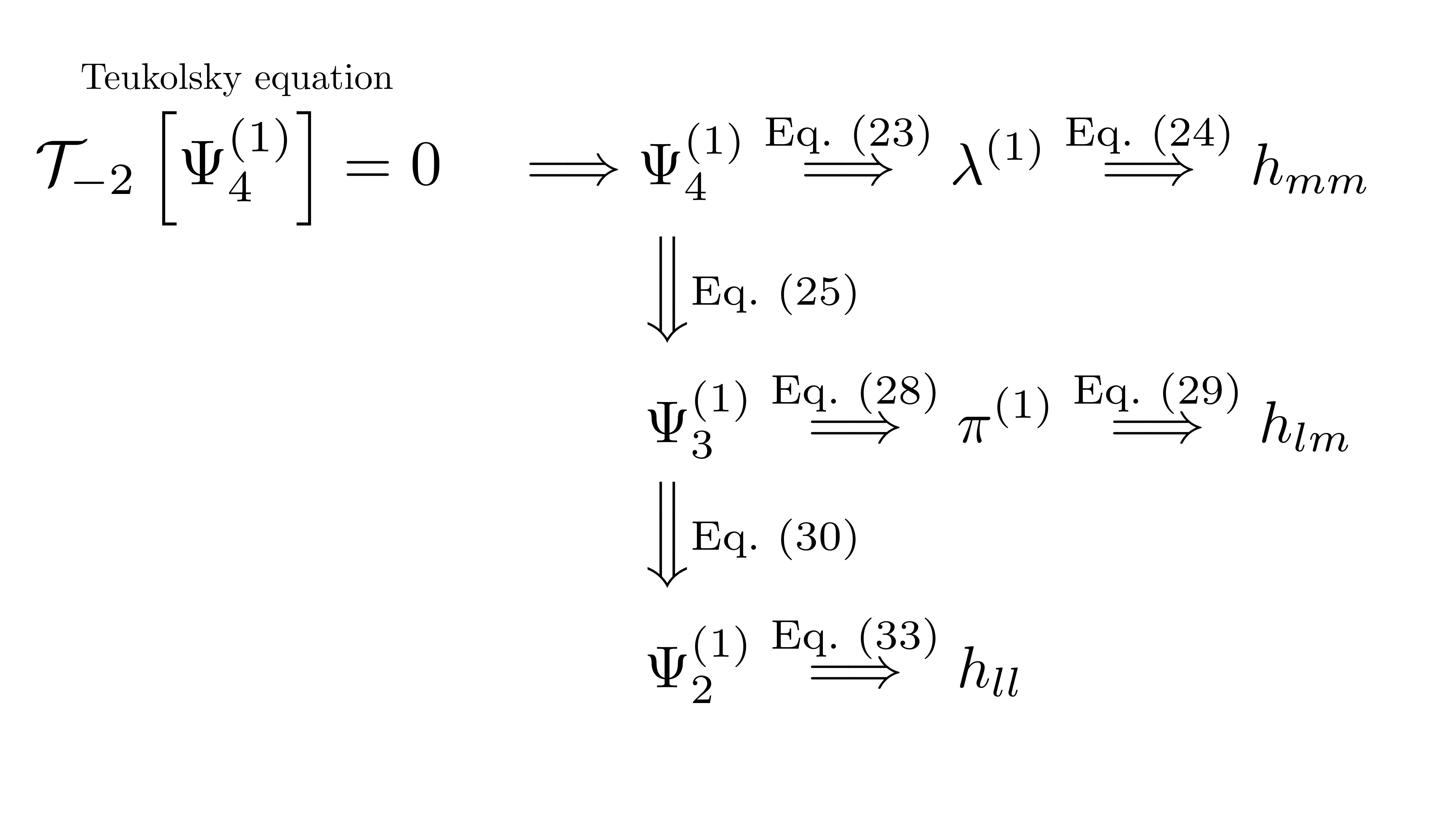}
\caption{\label{schematic}
Schematic of our procedure for metric reconstruction.
From the Teukolsky equation, one can solve for the
Weyl scalar $\Psi_{4}^{(1)}$.
In the outgoing radiation gauge detailed in
Sec.~\ref{section-gauge-metric}, one can then directly reconstruct the three
non-zero metric perturbations $h_{mm}$, $h_{lm}$,
and $h_{ll}$ using the Bianchi and Ricci identities
of the Newman-Penrose formalism.
}
\end{figure}

To preview the detailed derivation later in the paper,
in Fig.  ~\ref{schematic} we show a schematic of our
metric reconstruction procedure. In the outgoing radiation gauge,
the only non-zero metric perturbations $h_{\mu\nu}$
are the tetrad projections
$h_{mm}=h_{\mu\nu} m^\mu m^\nu, h_{lm}=h_{\mu\nu} l^\mu m^\nu$
and $h_{ll}=h_{\mu\nu} l^\mu l^\nu$, with the tetrad
consisting of a complex angular null vector $m^\mu$
and the real radially outgoing (ingoing) null vectors $l^\mu$ ($n^\mu$).
The starting point is to solve the Teukolsky equation
for the first order Weyl scalar $\Psi_{4}^{(1)}$.
One can then solve for the spin coefficient
$\lambda^{(1)}$ through Eq.~\eqref{eq:lambda-transport},
which can then be use to obtain $h_{mm}$
through Eq.~\eqref{eq:hmm-transport}.
Separately to this, one can obtain $\Psi_{3}^{(1)}$
from $\Psi_{4}^{(1)}$ using Eq.~\eqref{eq:psi-3-recon}.
The spin coefficient $\pi^{(1)}$ can then be obtained
from Eq.~\eqref{eq:pi-transport},
which then allows us to solve for $h_{lm}$
through Eq.~\eqref{eq:hlm-transport}.
Finally, from $\Psi_{3}^{(1)}$ we can obtain $\Psi_{2}^{(1)}$
from Eq.~\eqref{eq:psi-2-recon},
which in turn allows us to solve for $h_{ll}$ using
Eq.~\eqref{eq:h_ll-recon}.
The remaining first order spin coefficients can then be obtained
from Eqs.~\eqref{eq:lambda-1}-\eqref{eq:pi-1}
and the first order Weyl scalars
from Eqs.~\eqref{eq:psi_1-recon}-\eqref{eq:psi_0-recon}.

This kind of approach to metric reconstruction has a few advantages over the typical Hertz potential approach. First, using Hertz potentials requires one to work within one of the two radiation gauges, which place additional constraints on the matter sources, or need to be corrected via the method in ~\cite{Green:2019nam}. Here, though we have also chosen to work within the outgoing radiation gauge, this is simply because it is one of the easiest gauges to identify the necessary transport equations to fully reconstruct the metric. The basic strategy can be applied in essentially an arbitrary gauge, the only difference being the eventual number and complication of the transport equations to solve to obtain the first order metric. 
Second, the Hertz potentials are spin weight $\pm2$ quantities, and thus only have support for modes with $l \ge 2$. However, there are non-radiative modes with $l <2$ associated with shifts in the mass and spin of the black hole, and thus cannot be obtained from the Hertz potential. Our approach is able to re-construct these effects from homogeneous solutions to some of the transport equations, which we will detail in an upcoming paper. 
A third issue with the use of a Hertz potential is additional steps must be taken beyond simply 
applying Chrzanowski's operator if one needs the resultant metric to be consistent
with a desired $\Psi_4^{(1)}$. In particular, a fourth order null transport equation
needs to be solved; see e.g. Eq. (11) of \cite{Ori:2002uv}, and the discussion of its solution
therein.

The remainder of the paper is organized as follows.
In Sec.~\ref{chptr:NP_formalism} we list the
equations that govern perturbations of Type D spacetimes
to first and second order in perturbation theory,
a derivation of which is given in Appendix~\ref{section-master-equation}.
In Sec.~\ref{lin_met_and_guage} we
derive relations between first order NP quantities
and the linearized metric
(with the full list of expressions for the spin coefficients
given in Appendix~\ref{sec:linearized_np_scalars}),
and then describe the outgoing radiation gauge condition we use 
to fix the form of the first order metric perturbation.
In Sec.~\ref{met-recon} we describe our reconstruction 
procedure.
The path to go from $\Psi_4$ to $(h_{mm},h_{lm},h_{ll})$ described
there and 
illustrated in Fig.~\ref{schematic} is not unique,
and in Appendix~\ref{sec:alt_recon} we mention some alternative steps.
As an illustration,
in Sec.~\ref{case-study} we apply this method
to the case of quasi-normal modes of the Kerr spacetime in the limit
of spatial infinity, i.e. we expand about $r \rightarrow \infty$.
As explained in that section, there is a complication 
to finding the non-radiative metric perturbation associated
with changes in the mass and spin of the black hole due to the
gravitational wave perturbation; we leave it to future work
to address that issue.
In a companion paper~\cite{numerics_paper} we detail the numerical code that
implements the full method.
We conclude with a discussion of future work in Sec.~\ref{discuss}.
Throughout this work, we use units with $G = c =1$.
For the NP formalism, a brief review of which is given in
Appendix~\ref{sec:np_formalism},
we use the conventions of~\cite{Chandrasekhar_bh_book},
except that we use Greek letters to denote spacetime indices,
(e.g. our metric sign convention is $+---$, and we use 
$\bar{f}$ to denote the complex conjugate of $f$).

%%%%%%%%%%%%%%%%%%%%%%%%%%%%%%%%%%%%%%%%%%%%%%%%%%%%%%%%%%%%%%%%%%%%%%%%%%%%%%
%\section{Setup of basic formalism}
\section{Perturbations of Type D Spacetimes}
\label{chptr:NP_formalism}

In the non-spinning limit, perturbation theory can be performed at the
level of the metric, i.e. the metric can be written as
$
   g_{\mu \nu}
   =
   g_{\mu \nu}^{\rm Schw} + \zeta h_{\mu \nu} + {\cal{O}}(\zeta^{2})
   ,
$
   where $g_{\mu \nu}^{\rm Schw}$ is the background Schwarzschild metric,
$h_{\mu \nu}$ is the first order metric perturbation,
and $\zeta$ is an order keeping parameter.
One can then write out the field equations for $h_{\mu \nu}$,
which can be separated using
spin-weighted spherical harmonics~\cite{1967JMP.....8.2155G}.
The gravitational waves are then described by the Regge-Wheeler
(even parity)~\cite{PhysRev.108.1063}
and Zerilli (odd parity)~\cite{PhysRevLett.24.737, PhysRevD.2.2141} equations.
For Kerr black holes, and any generic type D spacetime,
the equations for the metric perturbation are not known to be separable.

%%%%%%%%%%%%%%%%%%%%%%%%%%%%%%%%%%%%%%%%%%%%%%%%%%%%%%%%%%%%%%%%%%%%%%%%%%%%%%%
%\subsection{Second order perturbations of Kerr}

The problem of finding separable equations for perturbations of Kerr spacetimes was
solved by Teukolsky using the NP formalism~\cite{Teukolsky:1973ha},
and Campanelli \& Lousto\cite{Campanelli:1998jv} extended this beyond
linear order. Here we list the equations, leaving a review of the
derivations to Appendix~\ref{section-master-equation}.
In the NP formalism, a gravitational wave perturbation is characterized
by the NP scalar $\Psi_4$ (or equivalently $\Psi_0$).
The equation for the linear vacuum perturbation
$\Psi_4^{(1)}$ is
\begin{align}
\label{eq:first_order_Teukolsky_main}
	\mathcal{T}\left[\Psi^{(1)}_4\right]
	= 0
	,
\end{align}
   where $\mathcal{T}$ is the Teukolsky operator for
a spin$=-2$ field (\ref{eq:Teukolsky_operator}). 
The equation for the second order vacuum perturbation
$\Psi_4^{(2)}$ is
\begin{align}
\label{eq:second_order_Teukolsky_main}
	\mathcal{T}\left[\Psi^{(2)}_4\right]
	=
	\mathcal{S}^{(2)}_{4}
	,
\end{align}
   where $\mathcal{T}$ is the same operator as in (\ref{eq:first_order_Teukolsky_main}), and
$\mathcal{S}_4^{(2)}$ is a second order ``source'' term:
\begin{widetext}
\begin{align}
\label{eq:second_order_Teukolsky_source_main}
	\mathcal{S}^{(2)}_{4} 
	\equiv &
-	\left[
		d_4^{(0)}\left(D+4\epsilon-\rho\right)^{(1)} 
	-	d_3^{(0)}\left(\delta+4\beta-\tau\right)^{(1)} 
	\right]
	\Psi_4^{(1)}
%	\nonumber \\
%	&
+	\left[
		d_4^{(0)}\left(\bar{\delta}+2\alpha+4\pi\right)^{(1)} 
	-	d_3^{(0)}\left(\Delta+2\gamma+4\mu\right)^{(1)} 
	\right]
	\Psi_3^{(1)}
	\nonumber \\
	&
-	3\left[
		d_4^{(0)}\lambda^{(1)}
	-	d_3^{(0)}\nu^{(1)}
	\right]
	\Psi^{(1)}_2
%	\nonumber \\
%	&
+	3\Psi^{(0)}_2\left[
		\left(d^{(1)}_4-3\mu^{(1)}\right)\lambda^{(1)}
	-	\left(d^{(1)}_3-3\pi^{(1)}\right)\nu^{(1)}
	\right]
	.
\end{align}
\end{widetext}
The source term is a function of first order perturbed 
NP spin coefficients
$\epsilon^{(1)},\rho^{(1)},\beta^{(1)},
\tau^{(1)},\alpha^{(1)},\pi^{(1)},\gamma^{(1)},\mu^{(1)},\lambda^{(1)},
\nu^{(1)}$, Weyl scalars $\Psi_{2}^{(1)},\Psi_{3}^{(1)},\Psi_{4}^{(1)}$,
and their derivatives through the background $d_3^{(0)},d_4^{(0)}$
and first order $D^{(1)},\Delta^{(1)},\delta^{(1)}$ gradient
operators
(see Appendix ~\ref{sec:np_formalism} and ~\ref{section-master-equation} 
for the relevant definitions).
This equation does not require imposing any particular coordinate
system on the background, although it does require using a background
tetrad that aligns with the two principal null directions of Kerr
(such as the Kinnersley tetrad).

We see that in this approach,
computing the leading nonlinear gravitational effects around
a Kerr black hole is reduced to computing the source term,
and then solving the Teukolsky equation with that source term.
If one has the first order metric perturbation
it is trivial to compute all the NP quantities
needed for the source term simply from their definitions.
However, what is more typical is to only have $\Psi_4^{(1)}$
from a solution to the first order Teukolsky equation.
As mentioned above then, the main technical challenge 
for the second order problem is reconstructing the
remaining NP quantities required for the source
from only one's knowledge of 
$\Psi_{4}^{(1)}$. In the remainder of this
paper we describe a method for doing so for
vacuum perturbations
(see~\cite{Green:2019nam} for a different
reconstruction procedure claimed to also work with 
gravity coupled to matter
that is smooth and of compact support).

\begin{comment}
Our strategy is the following:
\begin{enumerate}
\item By solving the first order Teukolsky equation \eqref{eq:first_order_Teukolsky}, we can control $\Psi_4^{(1)}$ from initial data, independently of the gauge. 
\item From the control obtained for $\Psi_4^{(1)}$ and after fixing a gauge, we develop a scheme to obtain control over all the metric coefficients and curvature components of the first order perturbation. Such obtained control allows us to compute the source term $\mathcal{S}^{(2)} $  in the second order Teukolsky equation \eqref{eq:second_order_Teukolsky_main}. 
\item We can therefore solve the second order Teukolsky equation \eqref{eq:second_order_Teukolsky_main} with known source and get control over the second order curvature perturbation $\Psi_4^{(2)}$. 
\end{enumerate}
\end{comment}

%%%%%%%%%%%%%%%%%%%%%%%%%%%%%%%%%%%%%%%%%%%%%%%%%%%%%%%%%%%%%%%%%%%%%%%%%%%%%%
\section{Linearized metric and gauge conditions}
\label{lin_met_and_guage}

Before describing our reconstruction procedure in the following section, here
we show the relation between linearized metric and tetrad components and linearized NP
scalars (Sec.\ref{lin_tetrad_our_gauge}), and then 
discuss the radiation gauge conditions we employ to fix
the form of the first order metric perturbation (Sec.~\ref{section-gauge-metric}).

\subsection{Linearized NP scalars 
in terms of the Linearized metric}
\label{lin_tetrad_our_gauge}

We write out the metric to first order in perturbation theory as $g_{\mu \nu} = g_{\mu \nu}^{B} + \zeta h_{\mu \nu} + {\cal{O}}(\zeta^{2})$, where $g_{\mu \nu}^{B}$ is a Petrov type D background spacetime, and $h_{\mu \nu}$ is the first order metric perturbation. For notational convenience, we write the components of $h_{\mu \nu}$ in the tetrad frame as $h_{ab} = h_{\mu \nu} e^{\mu}_{a} e^{\nu}_{b}$, reserving Latin (Greek) indices for tetrad (coordinate) components; for example $h_{nn} = h_{\mu \nu} n^{\mu} n^{\nu}$. We assume that the background tetrad $(l^{\mu}_{(0)}, n^{\mu}_{(0)}, m^{\mu}_{(0)},\bar{m}^{\mu}_{(0)})$ is chosen such that $\Psi_{0}^{(0)} = \Psi_{1}^{(0)} = \Psi_{3}^{(0)} = \Psi_{4}^{(0)} = \kappa^{(0)} = \sigma^{(0)} = \nu^{(0)} = \lambda^{(0)} = 0$.
Note that the results
in this subsection do not rely on the choice of gauge for the
metric, but do depend on the choice of the linearized tetrad.

Our starting point is to calculate the first order tetrad
in terms of the metric perturbation. 
The background tetrad forms a complete basis,
so it is natural to decompose the first order
tetrad in terms of these vectors, specifically
\begin{align}
\label{eq:tetrad-1}
	\begin{pmatrix} 
	l^{(1)}_{\mu}\\
	n^{(1)}_{\mu}\\
	m^{(1)}_{\mu}\\
	\bar{m}^{(1)}_{\mu}
	\end{pmatrix} 
	=
	\begin{pmatrix}
	b_{11}&b_{12}&c_{13}&\bar{c}_{13}\\
	b_{21}&b_{22}&c_{23}&\bar{c}_{23}\\
	c_{31}&c_{32}&c_{33}&c_{34}\\
	\bar{c}_{31}&\bar{c}_{32}&\bar{c}_{34}&\bar{c}_{33}
	\end{pmatrix}
	\begin{pmatrix} 
	l^{(0)}_{\mu}\\
	n^{(0)}_{\mu}\\
	m^{(0)}_{\mu}\\
	\bar{m}^{(0)}_{\mu}
	\end{pmatrix} 
	,
\end{align}
where the $b_{ij}$ are real coefficients and the $c_{ij}$ are complex coefficients. Following \cite{PhysRevD.13.806, Campanelli:1998jv}, we can use our six degrees of freedom for the linearized tetrad vectors to choose $b_{11}=c_{13}=c_{23}=\text{Im}c_{33}=0$. We now solve for the coefficients of the matrix in Eq.~\eqref{eq:tetrad-1} using the completeness relation $g_{\mu \nu} = 2 l_{(\mu} n_{\nu)} - 2 m_{(\mu} \bar{m}_{\nu)}$. Expanding to first order, we have
\begin{align}
h_{\mu \nu} = 2 l^{(1)}_{(\mu} n^{(0)}_{\nu)} + 2 l^{(0)}_{(\mu} n^{(1)}_{\nu)} - 2 m^{(1)}_{(\mu} \bar{m}^{(0)}_{\nu)} - 2 m^{(0)}_{(\mu} \bar{m}^{(1)}_{\nu)}\,.
\end{align}
Inserting the representation of the first order tetrad in Eq.~\eqref{eq:tetrad-1} and projecting into the tetrad frame gives us a set of linear equations that can be solved to obtain the $b$ and $c$ coefficients in terms of $h_{ab}$, specifically
\begin{subequations}
\begin{align}
\label{eq:l-1}
	l^{(1)}_{\mu}
	=&
	\frac{1}{2}h_{ll}n^{(0)}_{\mu}
	,\\
	n^{(1)}_{\mu}
	=&
	\frac{1}{2}h_{nn}l^{(0)}_{\mu}
+	h_{ln}n^{(0)}_{\mu}
	,\\
\label{eq:mbar-1}
	m^{(1)}_{\mu}
	=&
	h_{nm}l^{(0)}_{\mu}
+	h_{lm}n^{(0)}_{\mu}
-	\frac{1}{2}h_{m \om}m^{(0)}_{\mu}
-	\frac{1}{2}h_{mm}\bar{m}^{(0)}_{\mu}
	.
\end{align}
\end{subequations}
Raising the coordinate indices on these expressions involves flipping the signs of the $h_{ij}$ 
terms (since the relative signs of the covariant versus contravariant components of the
first order metric tensor perturbation are opposite). For convenience, we also write out the first order directional derivatives $(D,\Delta,\delta,\bar{\delta})$ using these relations:
\begin{subequations}
\begin{align}
\label{eq:D-1}
D^{(1)} &= -\frac{1}{2} h_{ll} \Delta^{(0)}\,,
\\
\label{eq:Delta-1}
\Delta^{(1)} &= -\frac{1}{2} h_{nn} D^{(0)} - h_{ln} \Delta^{(0)}\,,
\\
\label{eq:delta-1}
\delta^{(1)} &= - h_{nm} D^{(0)} - h_{lm} \Delta^{(0)} \nonumber\\
             &+ \frac{1}{2} h_{m \om} \delta^{(0)} + \frac{1}{2} h_{mm} \bar{\delta}^{(0)}\,.
\end{align}
\end{subequations}
The next step is to write out the spin coefficients in terms of the metric perturbations $h_{ab}$. To achieve this, we make use of the commutation relations in Eqs.~\eqref{eq:comm-1}-\eqref{eq:comm-4} and the first order tetrad in Eqs.~\eqref{eq:l-1}-\eqref{eq:mbar-1}. We expand out both sides of the commutation relations and match the coefficients of the directional derivatives to obtain linear equations for the first order spin coefficients. As an example of this, consider Eq.~\eqref{eq:comm-1}. Expanding out the left hand side, we have
\begin{widetext}
\begin{align}
\left[\delta, D\right]^{(1)} &= \frac{1}{2}\left[2 D^{(0)}h_{nm} + \left(\bar{\alpha}^{(0)} + \beta^{(0)} - \bar{\pi}^{(0)}\right)h_{m \om} + \left(\alpha^{(0)} + \bar{\beta}^{(0)} - \pi^{(0)}\right) h_{mm} 
\right.
\nn \\
&\left.
- 2\left(\gamma^{(0)} + \bar{\gamma}^{(0)}\right) h_{lm} + \bar{\nu}^{(0)} h_{ll} \right] D^{(0)} + \frac{1}{2}\left[2 D^{(0)} h_{lm} - \delta^{(0)}h_{ll} + \left(\bar{\alpha}^{(0)} + \beta^{(0)} - \tau^{(0)}\right) h_{ll} 
\right.
\nn \\
&\left.
- 2 \left(\epsilon^{(0)} + \bar{\epsilon}^{(0)}\right) h_{lm} + \kappa^{(0)} h_{m \om} + \bar{\kappa}^{(0)} h_{mm} \right] \Delta^{(0)} + \frac{1}{2} \left[-D^{(0)} h_{m \om} + \left(-\epsilon^{(0)} + \bar{\epsilon}^{(0)} - \bar{\rho}^{(0)}\right) h_{m\om} 
\right.
\nn \\
&\left.
- \left(-\gamma^{(0)} + \bar{\gamma}^{(0)} + \mu^{(0)}\right) h_{ll} - \bar{\sigma}^{(0)} h_{mm} + 2 \left(\pi^{(0)} + \bar{\tau}^{(0)}\right) h_{lm} \right] \delta^{(0)} 
\nn \\
&+ \frac{1}{2} \left[- D^{(0)} h_{mm} + \left(\epsilon^{(0)} - \bar{\epsilon}^{(0)} - \rho^{(0)}\right) h_{mm} - \bar{\lambda}^{(0)} h_{ll} - \sigma^{(0)} h_{m\om} + 2\left( \bar{\pi}^{(0)} + \tau^{(0)}\right) h_{lm}\right] \bar{\delta}^{(0)}\,.
\end{align}
Next, expanding out the right hand side, we obtain
\begin{align}
&\left[\bar{\alpha}^{(1)} + \beta^{(1)} - \bar{\pi}^{(1)} + \left(\epsilon^{(0)} - \bar{\epsilon}^{(0)}\right) h_{nm} - \frac{1}{2} \kappa^{(0)} h_{nn} + \bar{\rho}^{(0)} h_{nm} + \sigma^{(0)} h_{n \om}\right] D^{(0)}
\nn \\
&+ \left[\kappa^{(1)} - \frac{1}{2} \left(\bar{\alpha}^{(0)} + \beta^{(0)} - \bar{\pi}^{(0)}\right) h_{ll} + \left(\epsilon^{(0)} - \bar{\epsilon}^{(0)} + \bar{\rho}^{(0)}\right) h_{lm} - \kappa^{(0)} h_{ln} + \sigma^{(0)} h_{l \om}\right] \Delta^{(0)}
\nn \\
&+ \left[-\epsilon^{(1)} + \bar{\epsilon}^{(1)} - \bar{\rho}^{(1)} - \frac{1}{2} h_{m \om} \left(\epsilon^{(0)} - \bar{\epsilon}^{(0)} + \bar{\rho}^{(0)}\right) - \frac{1}{2} \sigma^{(0)} h_{\om \om}\right] \delta^{(0)}
\nn \\
&+ \left[-\sigma^{(1)} - \frac{1}{2} \left(\epsilon^{(0)} - \bar{\epsilon}^{(0)} + \bar{\rho}^{(0)}\right) h_{mm} - \frac{1}{2} \sigma^{(0)} h_{m \om}\right] \bar{\delta}^{(0)}
\end{align}
Matching the coefficients of $\Delta^{(0)}$ allows us to solve for $\kappa^{(1)}$, i.e.
\begin{align}
\label{eq:pert_ka_1}
\kappa^{(1)} &= \left(D - 2\epsilon - \bar{\rho}\right)^{(0)} h_{lm} - \frac{1}{2} \left(\delta - 2\bar{\alpha} - 2 \beta + \bar{\pi} + \tau\right)^{(0)} h_{ll}\,.
\end{align}
\end{widetext}

Repeating this method for the remaining commutation relations, we obtain
the rest of the linearized Newman-Penrose scalars written in terms
of the linearized metric components. We provide the complete listing of
these quantities in Appendix \ref{sec:linearized_np_scalars}.
The first order spin coefficients are now completely determined in terms of the metric perturbation.

The final step to complete the description in terms of the metric perturbation is to obtain the Weyl scalars. This can be done readily from the transport equations in Eqs.~\eqref{eq:riem-1}-\eqref{eq:riem-18}. As an example, we may obtain $\Psi_{0}^{(1)}$ directly from Eq.~\eqref{eq:riem-2}, 
due to the fact that $\sigma^{(0)} = 0 = \kappa^{(0)}$, specifically
\begin{align}
\label{eq:Psi_0-1}
\Psi_{0}^{(1)} &= \left(D - \rho - \bar{\rho} - 3\epsilon + \bar{\epsilon}\right)^{(0)} \sigma^{(1)} 
\nn \\
&- \left(\delta + \tau - \bar{\pi} + \bar{\alpha} + 3\beta\right)^{(0)} \kappa^{(1)}\,.
\end{align}
Likewise, from Eq.~\eqref{eq:riem-10}, we have
\begin{align}
\Psi_{4}^{(1)} &= \left(\bar{\delta} + 3\alpha + \bar{\beta} + \pi - \bar{\tau}\right)^{(0)} \nu^{(1)} 
\nn \\
&- \left(\Delta + \mu + \bar{\mu} + 3\gamma - \bar{\gamma}\right)^{(0)} \lambda^{(1)}\,.
\end{align}
The remaining Weyl scalars must be found by taking linear combinations of
Eqs.~\eqref{eq:riem-1}-\eqref{eq:riem-18}.
We here provide the exact representation of these without linearizing:
\begin{widetext}
\begin{subequations}
\begin{align}
\Psi_{1} &= \left(D - \bar{\rho} + \bar{\epsilon}\right)\beta - \left(\delta + \bar{\alpha} - \bar{\pi}\right)\epsilon - \left(\alpha + \pi\right)\sigma + \left(\mu + \gamma\right)\kappa\,,
\\
\Psi_{2} &= \frac{1}{3} \left[\left(\bar{\delta} - 2\alpha + \bar{\beta} - \pi - \bar{\tau}\right)\beta - \left(\delta - \bar{\alpha} + \bar{\pi} + \tau\right)\alpha + \left(D + \epsilon + \bar{\epsilon} + \rho - \bar{\rho}\right)\gamma - \left(\Delta - \bar{\gamma} - \gamma + \bar{\mu} - \mu\right)\epsilon 
\right.
\nn \\
&\left.
+ \left(\bar{\delta} - \alpha + \bar{\beta} - \bar{\tau} - \pi\right)\tau - \left(\Delta - \bar{\gamma} - \gamma + \bar{\mu} - \mu\right)\rho + 2\left(\nu \kappa - \lambda \sigma\right)\right]\,,
\\
\label{eq:Psi_3-1}
\Psi_{3} &= \left(\bar{\delta} + \bar{\beta} - \bar{\tau}\right)\gamma - \left(\Delta - \bar{\gamma} + \bar{\mu}\right)\alpha + \left(\rho + \epsilon\right)\nu - \left(\tau + \beta\right)\lambda\,.
\end{align}
\end{subequations}
\end{widetext}
This completes the description of NP quantities in terms of the metric perturbation.
%----------------------------------------------------------------------------------
%%%%%%%%%%%%%%%%%%%%%%%%%%%%%%%%%%%%%%%%%%%%%%%%%%%%%%%%%%%%%%%%%%%%%%%%%%%%%%
\subsection{Radiation gauges}
\label{section-gauge-metric}

As mentioned, the form of the Teukolsky equation given in the previous section is independent 
of the coordinate system, and only requires the radial null tetrad vectors 
to be aligned with the principle null directions of Kerr. Solving these equations in practice
requires choosing coordinates for the background metric and first order perturbations.
Here, we describe our gauge to fix the form of the first order metric and tetrad perturbations.

Under an infinitesimal gauge transformation $x^{\mu} \rightarrow x^{\mu} + \xi^{\mu}$ of the background metric, $h_{\mu \nu}$ transforms as
\begin{equation}
h_{\mu \nu} \rightarrow h_{\mu \nu} - \xi_{(\mu ; \nu)}\,.
\end{equation}
We make use of the radiation gauges developed by Chrzanowski~\cite{PhysRevD.11.2042}, in which
the metric perturbation is required to be transverse to one of the principal null directions. This condition can only be imposed in Type II spacetimes, or more symmetric spacetimes, like Type D~\cite{Price:2006ke}. For the outgoing radiation gauge, we begin by imposing
\begin{equation}
\label{eq:gauge-org}
n^{\mu} \left(h_{\mu \nu} - \xi_{(\mu ; \nu)}\right) = 0\,.
\end{equation}
This set of four equations for the vector $\xi^{\mu}$ imply we have freedom to choose $\xi^{\mu}$ such that four of the components of $h_{\mu \nu}$ are zero, specifically $h_{ln} = h_{nn} = h_{nm} = h_{n \om} = 0$ in this gauge.
However, in Petrov type D (or more generally Petrov type II)
spacetimes it turns out that we still have some residual gauge freedom
(related to the homogeneous solutions of Eq.~\eqref{eq:gauge-org})
that we can use to enforce a traceless condition~\cite{Price:2006ke}
\begin{equation}
\label{eq:gauge-trace}
{h_{\mu}}^{\mu} = g^{\mu \nu} h_{\mu \nu} = 0\,.
\end{equation}
Taken together with the previous conditions, this sets $h_{m \om} = 0$, leaving the
only nonzero components of the metric to be the real valued $h_{ll}$ and the complex valued $h_{lm}$ and $h_{mm}$. It then follows from Eqs.~\eqref{eq:nu-1},\eqref{eq:gamma-1}, \&~\eqref{eq:mu-1} that 
\begin{equation}
\label{eq:spin-org}
\nu^{(1)} = \mu^{(1)} = \gamma^{(1)} = 0\,.
\end{equation}
If coupling to matter, the traceless condition also imposes
a constraint on the stress energy tensor from Eq.~\eqref{eq:riem-14},
namely 
\begin{equation}
\label{eq:matter-org}
\Phi_{22} = 0 \qquad \Rightarrow \qquad T_{\mu \nu} n^{\mu} n^{\nu} = 0\,.
\end{equation}
Eqs.~\eqref{eq:gauge-org}-\eqref{eq:matter-org} specify the necessary and sufficient conditions for the outgoing radiation gauge. This gauge has the properties of being transverse and traceless on future null infinity and the past horizon for the Kerr spacetime.

Complementary to the outgoing radiation gauge, one can also specify the ingoing radiation gauge through the condition
\begin{equation}
l^{\mu} \left(h_{\mu \nu} - \xi_{(\mu ; \nu)}\right) = 0\,.
\end{equation}
Combining with the traceless condition in Eq.~\eqref{eq:gauge-trace}, we have the necessary conditions of the ingoing radiation gauge
\begin{align}
\label{eq:spin-irg}
\epsilon^{(1)} = \kappa^{(1)} &= \rho^{(1)} = 0\,,
\\
\label{eq:matter-irg}
\Phi_{00} = 0 \qquad &\Rightarrow \qquad T_{\mu \nu} l^{\mu} l^{\nu} = 0\,.
\end{align}
This gauge has the property of being transverse and traceless on past null infinity and the future null horizon of the Kerr spacetime. Either one of these gauges allow for metric reconstruction as outlined in this paper, so long as the matter stress energy tensor satisfies either Eq.~\eqref{eq:matter-org} or~\eqref{eq:matter-irg}. Since we are most interested in the problem of quasi-normal modes of Kerr black holes as the end state of a binary coalescence, we can restrict to the case of vacuum and both of these conditions are satisfied. For the remainder of this paper, we work within the outgoing radiation gauge.

%--------------------------------------------------------------------------------------
%%%%%%%%%%%%%%%%%%%%%%%%%%%%%%%%%%%%%%%%%%%%%%%%%%%%%%%%%%%%%%%%%%%%%%%%%%%%%%
\section{Reconstructing the metric from $\Psi_4^{(1)}$}
\label{met-recon}

In this section, we describe a procedure to reconstruct the
metric coefficients $h_{ll}$, $h_{l \om}$, and $h_{\om \om}$
in the outgoing radiation gauge from the Weyl curvature scalar $\Psi_4^{(1)}$.

In the NP formalism, there are eight complex equations from the Bianchi identities Eqs.~\eqref{eq:bianchi-1}-\eqref{eq:bianchi-8}, 36 complex equations (20 independent) from the Riemann identities Eqs.~\eqref{eq:riem-1}-\eqref{eq:riem-18}, and 12 complex equations for the spin coefficients Eqs.~\eqref{eq:lambda-1}-\eqref{eq:pi-1}. However, in our chosen gauge, we only need to solve for five real valued (one real and two complex) quantities. Thus, the problem of solving for the metric perturbation is overdetermined. The procedure that we detail below is, as a result, not unique, but it is sufficient to reconstruct the metric. Some alternative choices are outlined in Appendix~\ref{sec:alt_recon}.

To begin, we focus on solving for $h_{mm}$. Consider the Riemann identity in Eq.~\eqref{eq:riem-10}. This is one of the equations used to derive the Teukolsky equation, and as explained there, is already of first order smallness. Further, due to the choice of gauge, $\nu^{(1)} = 0$, and so we obtain the following transport equation for $\lambda^{(1)}$
\begin{equation}
\label{eq:lambda-transport}
\left(\Delta + \mu + \bar{\mu} + 3 \gamma - \bar{\gamma}\right)^{(0)} \lambda^{(1)} = - \Psi_{4}^{(1)}\,.
\end{equation}
Thus, once one has solved the Teukolsky equation for $\Psi_{4}^{(1)}$, one can naturally obtain $\lambda^{(1)}$. Now, consider the relationship between $\lambda^{(1)}$ and the metric perturbation in Eq.~\eqref{eq:lambda-1}. Once again, our choice of gauge eliminates all of the metric coefficients in this expression, except for $h_{\om \om}$. Thus, we obtain a transport equation for $h_{\om \om}$, namely
\begin{equation}
\label{eq:hmm-transport}
\left[\Delta + 2 \left(\bar{\gamma} - \gamma\right) + \bar{\mu} - \mu\right]^{(0)} h_{\om \om} = - 2\lambda^{(1)}\,.
\end{equation}
Of course, this also yields $h_{mm}$ since $h_{mm} = [h_{\om \om}]^{\dagger}$. The real and imaginary parts of $h_{mm}$ encode the gravitational waves at null infinity, and 
the above two equations are effectively equivalent to the statement
$\Psi_{4} = (1/2)\partial_t^2(h_{+} - i h_{\times})$
in a far field expansion, where $h_{+,\times}$ are the polarization states of gravitational waves. This
will become more explicit when we present our case study in Sec. \ref{case-study}.

Having solved for $h_{mm}$, we now turn our attention to $h_{lm}$. Consider the Bianchi identity in Eq.~\eqref{eq:bianchi-8}. Just like our starting point for $\lambda^{(1)}$, this equation was used to derive the Teukolsky equation and is already of first order smallness. Also, by virtue of $\nu^{(1)} = 0$, this gives us a transport equation that we may solve to obtain $\Psi_{3}^{(1)}$, namely
\begin{equation}
\label{eq:psi-3-recon}
\left(\Delta + 2\gamma + 4\mu\right)^{(0)} \Psi_{3}^{(1)} = \left(\delta - \tau + 4 \beta\right)^{(0)}\Psi_{4}^{(1)} + {\cal{R}}_{h}^{(1)}\,.
\end{equation}
For generality, we have kept the terms dependent on the Ricci scalars in the above equation. We will do so throughout the metric reconstruction procedure. However, these terms must satisfy the gauge condition in~\eqref{eq:matter-org}. Having solved for $\Psi_{3}^{(1)}$, we now consider the Riemann identity in Eq.~\eqref{eq:riem-9}. After linearizing, we have
\begin{align}
\label{eq:pi-transport-temp}
\left(\Delta + \gamma - \bar{\gamma}\right)^{(0)} \pi^{(1)} &= - \mu^{(0)} \left(\pi + \bar{\tau}\right)^{(1)} - \lambda^{(1)} \left(\bar{\pi} + \tau\right)^{(0)} 
\nn \\
&- \Psi_{3}^{(1)} - \Phi_{21}^{(1)}\,.
\end{align}
By combining Eqs.~\eqref{eq:pi-1} and the complex conjugate of Eq.~\eqref{eq:tau-1}, we find
\begin{equation}
\label{eq:tau+pi-1}
\pi^{(1)} + \bar{\tau}^{(1)} = - \frac{1}{2} h_{\om \om} \left(\bar{\pi} + \tau\right)^{(0)}\,.
\end{equation}
Combining this with Eq.~\eqref{eq:pi-transport-temp}, we obtain a transport equation for $\pi^{(1)}$,
\begin{align}
\label{eq:pi-transport}
\left(\Delta + \gamma - \bar{\gamma}\right)^{(0)} \pi^{(1)} &= \left(\frac{1}{2} \mu^{(0)} h_{\om \om} - \lambda^{(1)}\right) \left(\bar{\pi} + \tau\right)^{(0)} 
\nn \\
&- \Psi_{3}^{(1)} - \Phi_{21}^{(1)}\,.
\end{align}
Finally, by our choice of gauge, Eq.~\eqref{eq:pi-1} gives us the transport equation for $h_{l\om}$, namely
\begin{equation}
\label{eq:hlm-transport}
\left(\Delta + \bar{\mu} - 2\bar{\gamma}\right)^{(0)} h_{l\om} = -2 \pi^{(1)} - h_{\om \om}\tau^{(0)}\,.
\end{equation}
Once again, we can obtain $h_{lm}$ by taking the complex conjugate of $h_{l\om}$. Also, since we now have $h_{lm}$ and $h_{mm}$, we can directly calculate $\alpha^{(1)}$, $\beta^{(1)}$, and $\tau^{(1)}$ from Eqs.~\eqref{eq:alpha-1},\eqref{eq:beta-1}, and~\eqref{eq:tau-1}, respectively.

We now proceed with the final step and turn our attention to $h_{ll}$. Consider the Bianchi identity in Eq.~\eqref{eq:bianchi-7}. Linearizing, and applying our gauge conditions, we obtain a transport equation for the Weyl scalar $\Psi_{2}^{(1)}$,
\begin{equation}
\label{eq:psi-2-recon}
\left(\Delta + 3\mu\right)^{(0)}\Psi_{2}^{(1)} = \left(\delta + 2\beta -2\tau\right)^{(0)} \Psi_{3}^{(1)} + {\cal{R}}_{g}^{(1)}\,.
\end{equation}
Now consider the Riemann identity in Eq.~\eqref{eq:riem-6}, which after linearizing and applying gauge conditions becomes
\begin{align}
\label{eq:hll-eqn-temp}
D^{(1)} \gamma^{(0)} &+ \left(-\Delta+\gamma + \bar{\gamma}\right)^{(0)}\epsilon^{(1)} -  \gamma^{(0)} \left(\epsilon + \bar{\epsilon}\right)^{(1)}
\nn \\
&= \left(\alpha^{(1)} - \frac{1}{2} h_{\om \om} \beta^{(0)}\right) \left(\tau + \bar{\pi}\right)^{(0)} 
\nn \\
&+ \left(\beta^{(0)} - \frac{1}{2} h_{mm} \alpha^{(0)}\right) \left(\pi + \bar{\tau}\right)^{(0)}
\nn \\
& + \tau^{(1)} \pi^{(0)} + \tau^{(0)} \pi^{(1)} + \Psi_{2}^{(1)}\,,
\end{align}
where we have used Eq.~\eqref{eq:tau+pi-1}. The left hand side of this equation depends on $h_{ll}$ and its derivatives, while the right hand side is known from quantities already computed in the previous steps of metric reconstruction. Using Eq.~\eqref{eq:epsilon-1} and its complex conjugate, we have
\begin{align}
\epsilon^{(1)} + \bar{\epsilon}^{(1)} &= \frac{1}{2} \left(-\Delta + \gamma + \bar{\gamma}\right)^{(0)} h_{ll} - \left(\bar{\pi} + \tau\right)^{(0)} h_{l\om} 
\nn \\
&- \left(\pi + \bar{\tau}\right)^{(0)} h_{lm}\,.
\end{align}
Meanwhile, $D^{(1)}$ is given algebraically in terms of $h_{ll}$ through Eq.~\eqref{eq:D-1}. Combining these expressions with Eq.~\eqref{eq:hll-eqn-temp}, we obtain the following second order transport equation for $h_{ll}$
\allowdisplaybreaks[4]
\begin{widetext}
\begin{align}
\label{eq:h_ll-recon}
\Bigg[\frac{1}{4}\left(-\Delta + \gamma + \bar{\gamma}\right)^{(0)} &\left(-\Delta + 2\bar{\gamma} + \mu - \bar{\mu}\right)^{(0)} + \frac{1}{2} \gamma^{(0)} \left(-\Delta + \gamma + \bar{\gamma}\right)^{(0)} - \frac{1}{2} \Delta^{(0)} \gamma^{(0)}\Bigg] h_{ll}
\nn \\
&= \left[-\frac{1}{4} \left(-\Delta + \gamma + \bar{\gamma}\right)^{(0)} \left(-\delta + 2\bar{\alpha} - \bar{\pi} - 2\tau\right)^{(0)} + \gamma^{(0)} \left(\bar{\pi} + \tau\right)^{(0)}\right] h_{l\om}
\nn \\
&+ \left[-\frac{1}{4} \left(-\Delta + \gamma + \bar{\gamma}\right)^{(0)} \left(\bar{\delta} - 2\alpha - 3\pi - 2\bar{\tau}\right)^{(0)} + \gamma^{(0)} \left(\pi + \bar{\tau}\right)^{(0)}\right] h_{lm}
\nn \\
&+\left(\alpha^{(1)} - \frac{1}{2} \beta^{(0)} h_{\om \om}\right) \left(\bar{\pi} + \tau\right)^{(0)} + \left(\beta^{(1)} - \frac{1}{2} \alpha^{(0)} h_{mm}\right) \left(\pi + \bar{\tau}\right)^{(0)} 
\nn \\
&+ \pi^{(0)} \tau^{(1)} + \pi^{(1)} \tau^{(0)} + \Psi_{2}^{(1)}
\end{align}
\end{widetext}
Thus, we now have all of the necessary equations to solve for the components of the first order metric perturbation. The remaining spin coefficients and Weyl scalars not computed from the transport equations in this reconstruction procedure may be derived from these metric components through Eqs.~\eqref{eq:pert_ka_1}-\eqref{eq:pi-1} and Eqs.~\eqref{eq:Psi_0-1}-\eqref{eq:Psi_3-1}, respectively. In the next section, we give a practical example of this procedure.

%%%%%%%%%%%%%%%%%%%%%%%%%%%%%%%%%%%%%%%%%%%%%%%%%%%%%%%%%%%%%%%%%%%%%%%%%%%%%%
\section{Case Study: Quasi-normal modes of Kerr black holes}
\label{case-study}

Having developed a procedure to reconstruct the metric in the outgoing radiation gauge, we illustrate the method with a concrete example, namely the first order metric perturbation in the limit $r\rightarrow\infty$ corresponding to a single quasi-normal mode of a Kerr black hole. To address issues of mode coupling at second order will require reconstruction near the black hole, however this is sufficiently complicated that we will do so numerically, as described in the companion
paper~\cite{numerics_paper}. 

We work in Boyer-Lindquist coordinates
\begin{align}
ds^{2} &=  \left(1 - \frac{2Mr}{\Sigma}\right) dt^{2} + \frac{4Mra\sin^{2}\theta}{\Sigma}dt d\phi 
\nn \\
&- \frac{\Sigma}{\Delta} dr^{2} - \Sigma d\theta^{2} \nn \\
&- \left(r^{2} + a^{2} - \frac{2Mra^{2}}{\Sigma}\sin^{2}\theta\right) d\phi^{2}\,,
\end{align}
where $\Delta = r^{2} - 2Mr + a^{2}$,
and $\Sigma = r^{2} + a^{2} \cos^{2}\theta$,
and choose the Kinnersley tetrad~\cite{1969JMP....10.1195K},
(which sets $l^{\mu}$ and $n^{\mu}$ to be
parallel to the principal null directions of the Kerr spacetime): 
\begin{subequations}
\begin{align}
l^{\mu} &= \frac{1}{\Delta} \left( r^{2} + a^{2}, \Delta, 0 , a\right)\,,
\\
n^{\mu} &= \frac{1}{2\Sigma} \left(r^{2} + a^{2}, - \Delta, 0, a\right)\,,
\\
m^{\mu} &= \frac{1}{\sqrt{2} \; \Gamma} \left(i a \sin\theta, 0, 1, i \csc\theta\right)\,,
\end{align}
\end{subequations}
where $\Gamma = r + i a \cos\theta$. The spin coefficients 
and Weyl scalars are 
\begin{align}
\label{eq:kerr-sc}
   \kappa &= \sigma = \lambda = \nu = \epsilon 
   = \Psi_{0} = \Psi_{1} = \Psi_{3} = \Psi_{4} 
   = 0,
   \nonumber \\
   \rho &= - \frac{1}{\bar{\Gamma}},
   \qquad
   \beta = \frac{\cot\theta}{2^{3/2} \Gamma},
   \qquad
   \pi = \frac{i a \sin\theta}{2^{1/2} \bar{\Gamma}^{2}},
   \nonumber \\
   \tau &= - \frac{ia\sin\theta}{2^{1/2} \Gamma \bar{\Gamma}},
   \qquad
   \mu = - \frac{\Delta}{2 \Gamma \bar{\Gamma}^{2}},
   \qquad
   \gamma = \mu + \frac{r - M}{2\Gamma \bar{\Gamma}},
   \nonumber \\
   \alpha &= \pi - \bar{\beta},
   \qquad
   \Psi_{2} = - \frac{M}{\bar{\Gamma}^{3}}
   .
\end{align}
%
%------------------------------------------------------------------
\subsection{Solving the Teukolsky equation}
\label{teuk-sol}

Before we can reconstruct the metric, we need a solution for $\Psi_{4}^{(1)}$.
Teukolsky showed that by defining $\psi = \rho_{(0)}^{-4} \Psi_{4}^{(1)}$,
Eq.~\eqref{eq:first_order_Teukolsky}
can be solved by separation of variables~\cite{Teukolsky:1973ha};
we review that here.
In Boyer-Lindquist coordinates, and in vacuum
(i.e. all of the Ricci scalars are zero), the Teukolsky equation is
\begin{widetext}
\begin{align}
&\Bigg\{\left[\frac{(r^{2}+a^{2})}{\Delta} - a^{2} \sin^{2}\theta\right] \frac{\partial^{2}}{\partial t^{2}} + \frac{4Mar}{\Delta} \frac{\partial^{2}}{\partial t \partial \phi} - 4 \left[r + i a \cos\theta - \frac{M (r^{2} + a^{2})}{\Delta}\right] \frac{\partial}{\partial t} 
\nn \\
&- \Delta^{2} \frac{\partial}{\partial r} \left(\Delta^{-1} \frac{\partial}{\partial r}\right) - \frac{1}{\sin\theta} \frac{\partial}{\partial \theta} \left(\sin\theta \frac{\partial}{\partial\theta}\right) - \left(\frac{1}{\sin^{2}\theta} - \frac{a^{2}}{\Delta}\right) \frac{\partial^{2}}{\partial \phi^{2}} 
\nn \\
&+ 4 \left[\frac{a(r-M)}{\Delta} + \frac{i \cos\theta}{\sin^{2}\theta}\right] \frac{\partial}{\partial \phi} + \left(4 \cot^{2}\theta + 2\right)\Bigg\}\psi = 0\,.
\end{align}
By writing
\begin{equation}
\psi = e^{-i\omega t} e^{im\phi} R(r) S(\theta)\,,
\end{equation}
we can separate the above equation into
\begin{subequations}
\begin{align}
\label{eq:teuk-r}
\Delta^{2} \frac{d}{dr} \left(\Delta^{-1} \frac{dR}{dr}\right) + \left(\frac{K^{2} + 4i(r-M)K}{\Delta} - 8 i\omega r - B\right)R &= 0\,,
\\
\label{eq:teuk-th}
\frac{1}{s_{\theta}}\frac{d}{d\theta}\left(s_{\theta} \frac{dS}{d\theta}\right) + \left(a^{2} \omega^{2} c_{\theta}^{2} - \frac{m^{2}}{s_{\theta}^{2}} + 4 a \omega c_{\theta} + \frac{4 m c_{\theta}}{s_{\theta}^{2}} - \frac{4 c_{\theta}^{2}}{s_{\theta}^{2}}  - 2 + A\right) S &=0\,,
\end{align}
\end{subequations}
\end{widetext}
where $K = (r^{2} + a^{2}) \omega - a m$, $B = A + a^{2} \omega^{2} - 2 a m \omega$, $A = A_{lm}(a\omega)$ is a separation constant with eigenvalue $l$, and $(c_{\theta}, s_{\theta}) = (\cos\theta, \sin\theta)$. Eq.~\eqref{eq:teuk-th} provides the definition of spin-weighted spheroidal harmonics~\cite{10.2307/79369}, which reduce to the well known spin-weighted spherical harmonics in the limit $a\rightarrow 0$. We will write the solution to Eq.~\eqref{eq:teuk-th} as $S(\theta) = {_{-2}}S_{lm}(\theta)$.

To solve Eq.~\eqref{eq:teuk-r}, it is natural to make the transformation
\begin{equation}
Y = \frac{(r^{2} + a^{2})^{1/2}}{\Delta} R\,, \qquad \frac{dr_{\star}}{dr} = \frac{r^{2}+a^{2}}{\Delta}\,.
\end{equation}
Eq.~\eqref{eq:teuk-r} then reduces to
\begin{align}
\label{eq:Y-eq}
Y'' + \Big[&\frac{K^{2} + 4i(r-M) K - \Delta (8ir\omega + B)}{(r^{2}+a^{2})^{2}} 
\nn \\
&- G^{2} - G'\Big] Y = 0
\end{align}
where the prime corresponds to differentiation with respect to $r_{\star}$, and
\begin{equation}
G = \frac{r\Delta}{(r^{2}+a^{2})^{2}} - \frac{2(r-M)}{r^{2}+a^{2}}\,.
\end{equation}
We are interested in a solution near spatial infinity ($r\rightarrow \infty, r_{\star}\rightarrow \infty$); expanding 
in this limit, Eq.~\eqref{eq:Y-eq} becomes
\begin{equation}
Y'' + \left(\omega^{2} - \frac{4i\omega}{r}\right) Y = 0
\end{equation}
with solution $Y = (a_{0}/r^{2}) e^{-i\omega r_{\star}} + b_{0} r^{2}e^{i\omega r_{\star}}$. Since $(r^{2}+a^{2})^{1/2}/\Delta \sim 1/r$, this implies $R = (a_{0}/r) e^{-i\omega r_{\star}} + b_{0} r^{3} e^{i\omega r_{\star}}$. Transforming back to the original variable $\Psi_{4}^{(1)}$, we have
\begin{equation}
\Psi_{4}^{(1)} = \left(\frac{a_{0}}{r^{5}} e^{-i\omega r_{\star}} + \frac{b_{0}}{r} e^{i\omega r_{\star}}\right) e^{-i\omega t + i m \phi} {_{-2}}S_{lm}(\theta)\,.
\end{equation}
This solution corresponds to a superposition of ingoing ($e^{-i\omega r_{\star}}$) and outgoing ($e^{i\omega r_{\star}}$) radiation. To model the situation describing the ringdown of a black hole following a binary merger, we enforce the boundary condition that there is no ingoing radiation from infinity, i.e. $a_{0}=0$. Writing $b_{0} = _{-2}\!\!{\cal{A}}_{lm}$, we arrive at the desired asymptotic solution
\begin{equation}
\label{eq:psi-4-asym}
\Psi_{4}^{(1)} = \frac{_{-2}{\cal{A}}_{lm}}{r} e^{i\left[m\phi - \omega (t - r_{\star})\right]} {_{-2}}S_{lm}(\theta)\,.
\end{equation}
The complex constant $_{-2}{\cal{A}}_{lm}$ is determined by initial conditions, which in the case of a binary coalescence is determined by the inspiral and merger phases.

%------------------------------------------------------------------
\subsection{First order metric}
\label{firsto-met}

Having a solution for $\Psi_{4}^{(1)}$, we may now proceed to
reconstruct the first order metric perturbation.
Before we begin, there are a couple
of important points to mention, one related to 
modes for $l<2$, the other about inital data.
In general, perturbations of black holes
can have $l=0$ and $l=1$ angular modes,
which physically correspond to shifts in the mass $M$
and spin $a$ of the black hole. Such modes cannot
be captured by the spin $s=-2$ field $\Psi_4$ (or the $s=+2$ field
$\Psi_0$), since spin-weighted fields of spin $s$ can only have
support over $l\ge\rm max(|s|,|m|)$. 
As demonstrated in the companion paper~\cite{numerics_paper},
lack of knowledge of the $l=0,1$ modes does not affect
the source term or second order mode coupling 
from first order modes with $|m|\ge 2$. For radiative modes with $|m|<2$ the influence
of the non-radiative pieces need to be incorporated through 
a combination of non-trivial initial conditions for the transport equations,
and their homogeneous ($\Psi_{4}^{(1)}=0$) solutions,
which we leave to future work to investigate (for more
discussion of these issues see e.g. Appendix B of~\cite{,Andersson:2019dwi},
and~\cite{Dolan:2012jg} in the context of the self-force problem).
The example of metric reconstruction we provide here 
therefore does not include these non-radiative terms.

In regards to the specification of initial data, this is a non-trivial problem 
if posed on a spacelike (Cauchy) slice $\Sigma$, and, as with
the issues related to $l<2$ solutions of the transport
equations, we leave to future work to investigate. Though in brief,
the difficulty stems from the fact that initial data for the Einstein
equations (linearized or not) when posed on a spacelike hypersurface
is subject to the Hamiltonian and momentum constraints, most easily
expressed in terms of geometric objects and their gradients instrinsic to $\Sigma$.
In the NP formalism, only the two angular null tetrad vectors
$m$ and $\bar{m}$ can straight-forwardly be rotated
to be tangent to $\Sigma$ (see e.g.~\cite{Klainerman:2017nrb}); 
the other two null vectors, and more importantly the corresponding
gradient operators $D$ and $\Delta$ they define, contain pieces
orthogonal to $\Sigma$. Hence it is not easy to disentangle
what data is freely specifiable (beyond $\Psi_4^{(1)}$) versus constrained
if reconstruction is to begin on $\Sigma$. Here, the
imposition of the QNM ansatz for $\Psi_4^{(1)}$ for all time $t$, together with
only solving the equations in the large $r$ limit, skirts
the initial data issue\footnote{For the numerical solution discussed in the companion
paper we cannot make a QNM ansatz, and do
not limit the domain to large $r$. We still 
do not solve the initial data problem on $\Sigma$ there, but instead
circumvent the problem by a particular restriction of the class of initial data,
and only performing self consistent reconstruction within a related 
null wedge interior to the domain of the Cauchy evolution. Also, not all the NP equations
are used to reconstruct the metric, and a subset are
redundant (essentially stemming from the Bianchi identities).
These are used in the code to check that the reconstruction is in fact
self consistent within the null wedge. 
For details see \cite{Ripley:2020xby}}.

%To handle such modes, one must start with variations of the mass and
%spin of the Kerr metric and the effects that such variations
%have on NP quantities \cite{doi:10.1063/1.1666203}
%(for a more recent discussion see, e.g.
%Appendix B of~\cite{,Andersson:2019dwi}).
%Any variations in these quantities will depend on the
%specifics of the perturbations, but in theory will
%be related to the properties of infalling matter or
%radiation through the laws of black hole thermodynamics.
%Further, the modes with $l=0,1$ are known to contain
%non-radiative pieces in the self-force formalism~\cite{Dolan:2012jg}.

Our starting point will be to solve for $h_{mm}$. An intermediate step is to determine $\lambda^{(1)}$ from $\Psi_{4}^{(1)}$, with the relevant transport equation given in Eq.~\eqref{eq:lambda-transport}. We assume that $\lambda^{(1)}$ can be separated in a similar fashion to $\Psi_{4}^{(1)}$; specifically we write $\lambda^{(1)} = e^{-i\omega t} e^{im\phi} R_{\lambda}(r) S_{\lambda}(\theta)$, and our goal will be to determine $R_{\lambda}(r)$ and $S_{\lambda}(\theta)$. Inserting this ansatz into Eq.~\eqref{eq:lambda-transport}, applying the NP operators in Boyer-Lindquist coordinates, and expanding in $r \rightarrow \infty (r_{\star} \rightarrow \infty)$, we obtain
\begin{align}
- \frac{1}{2} e^{-i\omega t + i m \phi} &S_{\lambda}(\theta) \left(\frac{dR_{\lambda}}{dr} + i \omega R_{\lambda}(r)\right) 
\nn \\
&= - \frac{_{-2}{\cal{A}}_{lm}}{r} e^{-i\omega (t-r_{\star}) + i m\phi} {_{-2}}S_{lm}(\theta)\,.
\end{align}
A necessary condition to separate this equation is $S_{\lambda}(\theta) = {_{-2}}S_{lm}(\theta)$. Applying this, we obtain the following equation for $R_{\lambda}(r)$:
\begin{equation}
\label{eq:R_lambda}
\frac{dR_{\lambda}}{dr} + i \omega R_{\lambda} = - \frac{2}{r} {_{-2}}{\cal{A}}_{lm} e^{i\omega r_{\star}}
\end{equation}
The homogeneous solution to this equation scales as $e^{-i\omega r_{\star}}$, and thus the $(t,r_{\star})$ dependence of the full homogeneous solution goes as $\lambda^{(1)} \sim e^{-i\omega (t+r_{\star})}$. This corresponds to an ingoing mode, which we set to zero, and so we only need to worry about the particular solution to the above equation. Due to the behavior of the right hand side of Eq.~\eqref{eq:R_lambda}, the particular solution will scale as $e^{i\omega r_{\star}}$. Writing $R_{\lambda} = a_{0} e^{i\omega r_{\star}}/r^{n}$, we can insert this into Eq.~\eqref{eq:R_lambda}, and solve for $a_{0}$ and $n$ in an asymptotic expansion about spatial infinity. Doing so, we obtain $n=1$ and $a_{0} = - i {_{-2}}{\cal{A}}_{lm}/\omega$, and thus
\begin{equation}
\lambda^{(1)} = - \frac{i}{\omega r} {_{-2}}{\cal{A}}_{lm} e^{-i\omega (t-r_{\star}) + i m \phi} {_{-2}}S_{lm}(\theta)\,.
\end{equation}
Now that we have $\lambda^{(1)}$, we turn our attention to the transport equation for $h_{\om \om}$ given by Eq.~\eqref{eq:hmm-transport}. The procedure for determining $h_{\om \om}$ follows the same steps as finding $\lambda^{(1)}$. Writing $h_{\om \om} = e^{-i\omega t + i m\phi} R_{\om \om}(r) S_{\om \om}(\theta)$, the necessary condition for separability is $S_{\om \om}(\theta) = {_{-2}}S_{lm}(\theta)$. We then obtain the equation
\begin{equation}
\frac{dR_{\om \om}}{dr} + i \omega R_{\om \om} = - \frac{4 i}{\omega r} {_{-2}}{\cal{A}}_{lm} e^{i \omega r_{\star}}\,.
\end{equation}
Using our boundary condition to set the homogeneous solution to zero, we solve for the particular solution to obtain
\begin{equation}
h_{\om \om} = - \frac{2}{\omega^{2} r} {_{-2}} {\cal{A}}_{lm} e^{-i\omega (t-r_{\star}) + i m \phi} {_{-2}}S_{lm}(\theta)\,.
\end{equation}
Finally, taking the complex conjugate, we have
\begin{equation}
h_{mm} = - \frac{2}{\bar{\omega}^{2} r} {_{-2}}\bar{{\cal{A}}}_{lm} e^{i \ow (t-r_{\star}) - i m \phi} {_{-2}}\bar{S}_{lm}(\theta)\,.
\end{equation}
We have made it explicit here that one has to take the complex conjugate of $\omega$ and ${_{-2}}S_{lm}$, as well as ${_{-2}}{\cal{A}}_{lm}$. In general, the frequency of the quasi-normal modes is complex, and since ${_{-2}}S_{lm}$ depends on $\omega$, then it is also complex.

We now turn our attention to solving for $h_{lm}$. The starting point is to solve for the Weyl scalar $\Psi_{3}^{(1)}$ from Eq.~\eqref{eq:psi-3-recon}. Expanding the right hand side of this equation, we obtain
\begin{align}
\left(\delta - \tau + 4 \beta\right)^{(0)} \Psi_{4}^{(1)} &= \frac{{_{-2}}{\cal{A}}_{lm}}{\sqrt{2} r^{2}} e^{im\phi - i \omega (t-r_{\star})} 
\nn \\
&\times {\cal{L}}_{-2} \left[{_{-2}}S_{lm}(\theta)\right]\,,
\end{align}
where ${\cal{L}}_{s} = \partial_{\theta} - m\csc\theta - s \cot\theta + a\omega \sin\theta$. These are the same operators that appear in the well-known Teukolsky-Starobinsky identities~\cite{Chandrasekhar_bh_book}. It is worth pointing out, however, that the operation ${\cal{L}}_{-2}[{_{-2}}S_{lm}(\theta)]$ does not generate the spin-weight $-1$ spheroidal harmonic ${_{-1}}S_{lm}(\theta)$, which can be verified by direct application of the angular Teukolsky equation~\eqref{eq:teuk-th} for spin-weight $-1$. In fact, this is the reason why one cannot decouple the equations governing electromagnetic and gravitational perturbations of the Kerr-Newman spacetime~\cite{Giorgi:2020ujd, Chandrasekhar_bh_book}. Note that, in the non-spinning limit (i.e. $a=0$), the operator ${\cal{L}}_{s}$ does reduce to the raising operator for spin-weighted spherical harmonics ${_{s}}Y_{lm}(\theta, \phi)$, and in the Geroch-Held-Penrose formalism~\cite{GHP_paper}, is the asymptotically expanded $\edth$ operator which raises the spin-weight of quantities. Thus, we may expect that the operation ${\cal{L}}_{-2}[{_{-2}}S_{lm}(\theta)]$ does produce an angular function of spin-weight $-1$, but that it does not satisfy the corresponding angular Teukolsky equation.

To solve for $\Psi_{3}^{(1)}$, we propose the ansatz $\Psi_{3}^{(1)} = e^{im\phi - i\omega t} R_{3}(r) S_{3}(\theta)$. In order to perform separation of variables, we must have $S_{3}(\theta) = {\cal{L}}_{-2}[{_{-2}}S_{lm}(\theta)]$. This gives us the equation for the radial function $R_{3}(r)$ in the limit $r \rightarrow \infty$
\begin{align}
\frac{dR_{3}}{dr} + i \omega R_{3}(r) = \frac{{_{-2}}{\cal{A}}_{lm}}{\sqrt{2} r^{2}} e^{i\omega r_{\star}} 
\end{align}
Solving this equation, we obtain
\begin{equation}
\Psi_{3}^{(1)} = \frac{i}{\sqrt{2}} \frac{{_{-2}}{\cal{A}}_{lm}}{\omega r^{2}} e^{im\phi - i\omega (t-r_{\star})} {\cal{L}}_{-2}\left[{_{-2}}S_{lm}(\theta)\right]\,.
\end{equation}
The remainder of the procedure to obtain $\pi^{(1)}$ and $h_{lm}$ follows these exact same steps. The angular dependence of these functions is ${\cal{L}}_{-2}[{_{-2}}S_{lm}(\theta)]$ in order to perform separation of variables. The end result of this computation is
\begin{subequations}
\begin{align}
\pi^{(1)} &= \frac{1}{\sqrt{2}} \frac{{_{-2}}{\cal{A}}_{lm}}{\omega^{2} r^{2}} e^{im\phi - i\omega (t-r_{\star})} {\cal{L}}_{-2}\left[{_{-2}}S_{lm}(\theta)\right]\,,
\\
h_{l\om} &= -i\sqrt{2} \frac{{_{-2}}{\cal{A}}_{lm}}{\omega^{3} r^{2}} e^{im\phi - i\omega (t-r_{\star})} {\cal{L}}_{-2}\left[{_{-2}}S_{lm}(\theta)\right]\,.
\end{align}
\end{subequations}
By virtue of having solved for $\pi^{(1)}$ and $h_{lm}$, we may also compute $\alpha^{(1)}$, $\beta^{(1)}$, and $\tau^{(1)}$, with the end result being
\begin{subequations}
\begin{align}
\alpha^{(1)} &= \frac{1}{2^{3/2}} \frac{{_{-2}}{\cal{A}}_{lm}}{\omega^{2} r^{2}} e^{im\phi - i \omega (t-r_{\star})} 
\nn \\
&\times \left\{2 {\cal{L}}_{-2}\left[{_{-2}}S_{lm}(\theta)\right] - \cot\theta {_{-2}}S_{lm}(\theta)\right\}\,,
\\
\beta^{(1)} &= \frac{1}{2^{3/2}} \frac{{_{-2}}\bar{{\cal{A}}}_{lm}}{\bar{\omega}^{2} r^{2}} e^{-im\phi + i \bar{\omega} (t-r_{\star})} \cot\theta {_{-2}}\bar{S}_{lm}(\theta)\,,
\\
\tau^{(1)} &= - \frac{1}{\sqrt{2}} \frac{{_{-2}}\bar{\cal{A}}_{lm}}{\bar{\omega}^{2} r^{2}} \left\{{\cal{L}}_{-2}\left[{_{-2}}S_{lm}(\theta)\right]\right\}^{\dagger}\,,
\end{align}
\end{subequations}
where $\dagger$ corresponds to complex conjugation of the angular function.

Finally, we consider the solution for $h_{ll}$. The first step is to solve for $\Psi_{2}^{(1)}$ using Eq.~\eqref{eq:psi-2-recon}. Expanding the right hand side, we have
\begin{equation}
\left(\delta + 2\beta - 2\tau\right)^{(0)}\Psi_{3}^{(1)} = \frac{i}{2} \frac{{_{-2}}{\cal{A}}_{lm}}{\omega r^{3}} {\cal{L}}_{-1} {\cal{L}}_{-2} \left[{_{-2}}S_{lm}(\theta)\right]\,.
\end{equation}
Writing the ansatz $\Psi_{2}^{(1)} = e^{im\phi - i \omega t} R_{2}(r) S_{2}(\theta)$, and expanding the left hand side of Eq.~\eqref{eq:psi-2-recon}, we have that $S_{2}(\theta) = {\cal{L}}_{-1} {\cal{L}}_{-2}[{_{-2}}S_{lm}(\theta)]$ in order to achieve separation of variables. We are then left with
\begin{equation}
\frac{dR_{2}}{dr} + i\omega R_{2}(r) = \frac{i}{2} \frac{{_{-2}}{\cal{A}}_{lm}}{\omega r^{3}} e^{i\omega r_{\star}}\,,
\end{equation}
which can be solved in a $1/r$ expansion to obtain
\begin{equation}
\Psi_{2}^{(1)} = - \frac{1}{2} \frac{{_{-2}}{\cal{A}}_{lm}}{\omega^{2} r^{3}} e^{im\phi - i \omega (t-r_{\star})} {\cal{L}}_{-1} {\cal{L}}_{-2}\left[{_{-2}}S_{lm}(\theta)\right]\,.
\end{equation}

With $\Psi_{2}^{(1)}$ in hand, we now turn to Eq.~\eqref{eq:h_ll-recon}. Consider the source terms on the right hand side of this equation. In an $r\rightarrow\infty$ expansion, the terms containing $h_{l\om}$, $h_{lm}$, and $\Psi_{2}^{(1)}$ dominate, and scale as $1/r^{3}$. This expanded source term is real valued, since $h_{ll}$ must be real valued. Writing $h_{ll} = e^{im\phi - i \omega t} R_{+}(r) S_{+}(\theta) + e^{-im\phi + i\omega t} R_{-}(r) S_{-}(\theta)$, the necessary conditions to perform separation of variables are $S_{+}(\theta) = {\cal{L}}_{-1}{\cal{L}}_{-2}[{_{-2}}S_{lm}(\theta)]$ and $S_{-}(\theta) = \{ {\cal{L}}_{-1} {\cal{L}}_{-2}[{_{-2}}S_{lm}(\theta)]\}^{\dagger}$. Expanding about $r\rightarrow\infty$, we obtain
\begin{subequations}
\begin{align}
\frac{d^{2} R_{+}}{dr^{2}} + 2i \omega \frac{dR_{+}}{dr} - \omega^{2} R_{+}(r) &= - 4\frac{{_{-2}}{\cal{A}}_{lm}}{\omega^{2} r^{3}} e^{i \omega r_{\star}}\,,
\\
\frac{d^{2} R_{-}}{dr^{2}} - 2i \bar{\omega} \frac{dR_{-}}{dr} - \bar{\omega}^{2} R_{-}(r) &= -4 \frac{{_{-2}}\bar{\cal{A}}_{lm}}{\bar{\omega}^{2} r^{3}} e^{-i\bar{\omega} \!\! r_{\star}}\,.
\end{align}
\end{subequations}
These equations can be solved with the methods we have previously employed to obtain
\begin{align}
h_{ll} &= 4 \frac{{_{-2}}{\cal{A}}_{lm}}{\omega^{4} r^{3}} e^{im\phi - i\omega(t-r_{\star})} {\cal{L}}_{-1} {\cal{L}}_{-2}\left[{_{-2}}S_{lm}(\theta)\right] + \text{c.c}\,,
\end{align}
where c.c. is shorthand for the complex conjugate of the preceding term.

Now that we have all of the components of the metric in our chosen gauge, we may complete the first order description of the NP quantities. Applying Eqs.~\eqref{eq:pert_ka_1}-\eqref{eq:pi-1}, the remaining spin coefficients are
\begin{subequations}
\begin{align}
\kappa^{(1)} &= -\sqrt{2} i \frac{{_{-2}}\bar{\cal{A}}_{lm}}{\bar{\omega}^{3} r^{3}} e^{-im\phi + i\ow(t-r_{\star})} \left\{{\cal{L}}_{-2}\left[{_{-2}}S_{lm}(\theta)\right]\right\}^{\dagger}\,,
\\
\sigma^{(1)} &= \frac{{_{-2}}\bar{\cal{A}}_{lm}}{\bar{\omega}^{2} r^{2}} e^{-im\phi + i \ow (t-r_{\star})} {_{-2}}\bar{S}_{lm}(\theta)\,,
\\
\epsilon^{(1)} &= \frac{5i}{4} \frac{{_{-2}}{\cal{A}}_{lm}}{\omega^{3} r^{3}} e^{im\phi - i \omega (t-r_{\star})} {\cal{L}}_{-1} {\cal{L}}_{-2}\left[{_{-2}}S_{lm}(\theta)\right]
\nn \\
& - \frac{3i}{4} \frac{{_{-2}}\bar{\cal{A}}_{lm}}{\bar{\omega}^{3} r^{3}} e^{-im\phi + i\ow (t-r_{\star})} \left\{{\cal{L}}_{-1} {\cal{L}}_{-2}\left[{_{-2}}S_{lm}(\theta)\right]\right\}^{\dagger}\,,
\\
\rho^{(1)} &= \frac{i}{2} \frac{{_{-2}}{\cal{A}}_{lm}}{\omega^{3} r^{3}} e^{im\phi - i\omega(t-r_{\star})} {\cal{L}}_{-1} {\cal{L}}_{-2}\left[{_{-2}}S_{lm}(\theta)\right] + \text{c.c}\,.
\end{align}
\end{subequations}
To obtain the remaining Weyl scalar $\Psi_{1}^{(1)}$ and $\Psi_{0}^{(1)}$, we use the linearize Bianchi identities in Eqs.~\eqref{eq:Psi_1-eqn} \&~\eqref{eq:Psi_0-eqn}. The methods for solving these are the exact same methods we detailed for the metric coefficients. The end result is
\begin{subequations}
\begin{align}
\label{eq:Psi_1-eqn}
\Psi_{1}^{(1)} &= \frac{i}{\sqrt{2}} \frac{{_{-2}}{\cal{A}}_{lm}}{\omega^{3} r^{4}} e^{im\phi - i\omega (t-r_{\star})} {\cal{L}}_{0} {\cal{L}}_{-1} {\cal{L}}_{-2}\left[{_{-2}}S_{lm}(\theta)\right],
\\
\label{eq:Psi_0-eqn}
\Psi_{0}^{(1)} &= \frac{{_{-2}}{\cal{A}}_{lm}}{\omega^{4} r^{5}} e^{im\phi - i\omega (t-r_{\star})} {\cal{L}}_{1} {\cal{L}}_{0} {\cal{L}}_{-1} {\cal{L}}_{-2} \left[{_{-2}}S_{lm}(\theta)\right]
\nn \\
&-6iM \frac{{_{-2}}\bar{\cal{A}}_{lm}}{\bar{\omega}^{3} r^{5}} e^{-im\phi + i \ow (t-r_{\star})} {_{-2}}\bar{S}_{lm}(\theta)\,.
\end{align}
\end{subequations}
This completes the derivation of all NP quantities at first order.
%%%%%%%%%%%%%%%%%%%%%%%%%%%%%%%%%%%%%%%%%%%%%%%%%%%%%%%%%%%%%%%%%%%%%%%%%%%%%%
\section{Discussion}
\label{discuss}

Here we have laid some of the ground work necessary for the study
of second order perturbations of Kerr black holes.
Working in outgoing radiation gauge,
we showed that the first order metric perturbations of a Kerr
black hole can be reconstructed starting from a single NP quantity,
namely $\Psi_{4}^{(1)}$.
As an example we have applied this to obtain the first order metric perturbations
associated with the quasi-normal modes of Kerr black holes in the
asymptotic limit.

There are several directions for future work.
As mentioned, reconstructing the metric over the entire
spacetime is complicated, and might not be analytically tractable.
We have developed a numerical code to implement 
the solution of the Teukolsky equation, and reconstruction procedure,
over the full spacetime exterior to the 
horizon~\cite{numerics_paper}. 
This is particularly relevant regarding questions of mode-coupling
after binary black hole mergers, as this 
phenomena will be governed 
by sources strongest in the near horizon region.
Another direction of future study would thus be to investigate 
whether, in addition to our
numerical analysis, analytic solutions may be obtained there. 
Also as discussed in Sec.~\ref{firsto-met}, additional
work is needed to solve for corrections to the metric
corresponding to changes in the spin and mass of the black hole.

As mentioned in the introduction, crucial to understanding
the nonlinear regime of ringdown is the question of what the
``initial conditions'' of the perturbed black hole 
following a merger are. If this is not known,
it would be difficult to distinguish the higher overtones of linear
modes from second order effects,
which could have similar amplitudes,
frequencies and decay rates\footnote{If--as argued
in~\cite{Giesler:2019uxc}--linear theory can very
accurately describe post merger ringdown
dynamics from peak amplitude onward,
second order analysis presumably then should
be able to extend this to some time {\em before} peak amplitude.}. 
The close limit approximation to black hole
mergers~\cite{Price:1994pm} seems like a natural avenue to address
the question of initial conditions.
Insight could also be gained from recent studies investigating this in the 
EMRI limit~\cite{Apte:2019txp,Lim:2019xrb}.
Also, numerical simulations of mergers can be used
to at least constrain the initial conditions
via measurement of ``final conditions'', i.e.,
the amplitudes and phases of modes in the ringdown
once all the nonlinear effects have sorted themselves out,
as well as measure driven second order modes
that will persist and look like QNMs with amplitudes and complex frequencies 
that are squares of their parent modes (see e.g.~\cite{London:2014cma}).

A further interesting application is investigating
the energy cascade between modes
due to nonlinear effects in ringdown.
In asymptotically Anti de-Sitter (AdS) spacetime,
several studies of black holes and black branes have shown
that horizon perturbations, modulo the natural decay,
become turbulent~\cite{Carrasco:2012nf,Green:2013zba,Adams:2013vsa}.
This may be a peculiarity of AdS spacetime,
though a study in~\cite{2015PhRvL.114h1101Y} suggested similar
phenomenology might be present for
very rapidly rotating Kerr black holes in
asymptotically flat spacetime. Those researchers used a
scalar field on a Kerr background as a model for
gravitational wave perturbations; with the tools
presented here and in~\cite{numerics_paper}
it should be possible to repeat this for tensor perturbations.
Their work suggests that turbulent dynamics might only be apparent
for very rapidly spinning black holes;
whether these exist in nature is unknown,
nevertheless this is still an interesting open theoretical problem.

\acknowledgements

We would like to thank Andrew Spiers for aid in checking the first order spin coefficients in Eqs.~\eqref{eq:lambda-1}-\eqref{eq:pi-1}. N.L. \& F.P. acknowledge support from NSF grant PHY-1912171, the Simons Foundation, and the Canadian Institute for Advanced Research (CIFAR). E.G. acknowledges support from NSF grant DMS-2006741.
%%%%%%%%%%%%%%%%%%%%%%%%%%%%%%%%%%%%%%%%%%%%%%%%%%%%%%%%%%%%%%%%%%%%%%%%%%%%%%
%%%%%%%%%%%%%%%%%%%%%%%%%%%%%%%%%%%%%%%%%%%%%%%%%%%%%%%%%%%%%%%%%%%%%%%%%%%%%%
\appendix
%%%%%%%%%%%%%%%%%%%%%%%%%%%%%%%%%%%%%%%%%%%%%%%%%%%%%%%%%%%%%%%%%%%%%%%%%%%%%%
\section{Newman-Penrose formalism}
\label{sec:np_formalism}
   For completeness in this Appendix we
review the Newman-Penrose (NP) formalism.
We use the conventions of~\cite{Chandrasekhar_bh_book},
(e.g. our metric sign convention is $+---$, and we use 
$\bar{f}$ to denote the complex conjugate of $f$),
except that we use Greek letters to denote spacetime indices,

The NP formalism is a re-formulation of the Einstein field equations in a null tetrad frame, defined by four null vectors $e^{\mu}_{a} = (l^{\mu}, n^{\mu}, m^{\mu}, \bar{m}^{\mu})$ satisfying
\begin{equation}
\label{eq:tetrad-dot}
l^{\mu} n_{\mu} = 1\,, \qquad m^{\mu} \bar{m}_{\mu} = -1
\end{equation}
where the over-bar corresponds to complex conjugation, and the remaining dot products are zero. The metric $g_{\mu \nu}$ is related to the null vectors via $g_{\mu \nu} = \eta_{ab} e^{a}_{\mu} e^{b}_{\nu}$, where 
\begin{equation}
\eta_{ab} = \begin{bmatrix}
		0 & 1 & 0 & 0\\
		1 & 0 & 0 & 0\\
		0 & 0 & 0 & -1\\
		0 & 0 & -1 & 0
	\end{bmatrix}\,.
\end{equation}
This leads to the completeness relation
\begin{equation}
\label{eq:complete}
g_{\mu\nu} = 2 l_{(\mu} n_{\nu)} - 2 m_{(\mu} \bar{m}_{\nu)}
\end{equation}
We further define the derivatives along the null directions as
\begin{align}
{D} &= l^{\mu} \partial_{\mu}\,, \qquad {\Delta} = n^{\mu} \partial_{\mu}\,,
\nn \\
{\delta} &= m^{\mu} \partial_{\mu}\,, \qquad \bar{{\delta}} = \bar{m}^{\mu} \partial_{\mu}\,.
\end{align}
These differential operators satisfy the following commutation relations
\begin{subequations}
\begin{align}
\label{eq:comm-1}
[\Delta, D] &= (\gamma + \bar{\gamma})D + (\epsilon + \bar{\epsilon}) \Delta - (\bar{\tau} + \pi) \delta 
\nn \\
&- (\tau +\bar{\pi}) \bar{\delta}\,,
\\
\label{eq:comm-2}
[\delta, D] &= (\bar{\alpha} + \beta - \bar{\pi}) D + \kappa \Delta - (\bar{\rho} + \epsilon - \bar{\epsilon})\delta - \sigma \bar{\delta}\,,
\\
\label{eq:comm-3}
[\delta, \Delta] &= - \bar{\nu} D + (\tau-\bar{\alpha}-\beta)\Delta + (\mu-\gamma+\bar{\gamma})\delta + \bar{\lambda}\bar{\delta}\,,
\\
\label{eq:comm-4}
[\bar{\delta}, \delta] &= (\bar{\mu} - \mu) D + (\bar{\rho}-\rho) \Delta + (\alpha - \bar{\beta}) \delta 
\nn \\
&+ (\beta - \bar{\alpha}) \bar{\delta}\,,
\end{align}
\end{subequations}
where $\{\alpha, \beta, \gamma, \epsilon, \rho, \lambda, \pi, \mu, \nu, \tau, \sigma, \kappa\}$ are the complex spin coefficients. The components of curvature in the NP formalism are characterized by contractions of the null tetrad with the Weyl tensor and Ricci tensor; specifically, the Weyl tensor contractions are
\begin{subequations}
\begin{align}
\Psi_{0} &= - C_{\mu \nu \rho \sigma} l^{\mu} m^{\nu} l^{\rho} m^{\sigma}\,,
\\
\Psi_{1} &= - C_{\mu \nu \rho \sigma} l^{\mu} n^{\nu} l^{\rho} m^{\sigma}\,,
\\
\Psi_{2} &= - C_{\mu \nu \rho \sigma} l^{\mu} m^{\nu} \bar{m}^{\rho} n^{\sigma}\,,
\\
\Psi_{3} &= - C_{\mu \nu \rho \sigma} l^{\mu} n^{\nu} \bar{m}^{\rho} n^{\sigma}\,,
\\
\Psi_{4} &= - C_{\mu \nu \rho \sigma} n^{\mu} \bar{m}^{\nu} n^{\rho} \bar{m}^{\sigma}\,,
\end{align}
\end{subequations}
and the contractions with the Ricci tensor are
\begin{subequations}
\begin{align}
\label{eq:phi_00}
\Phi_{00} &= - \frac{1}{2} R_{\mu \nu} l^{\mu} l^{\nu}\,, \qquad \Phi_{22} = - \frac{1}{2} R_{\mu \nu} n^{\mu} n^{\nu}\,,
\\
\Phi_{02} &= -\frac{1}{2} R_{\mu \nu} m^{\mu} m^{\nu}\,, \qquad \Phi_{20} = - \frac{1}{2} R_{\mu \nu} \bar{m}^{\mu} \bar{m}^{\nu}
\\
\Phi_{11} &= -\frac{1}{4} R_{\mu \nu} \left(l^{\mu} n^{\nu} + m^{\mu} \bar{m}^{\nu}\right)\,,
\\
\Phi_{01} &= - \frac{1}{2} R_{\mu \nu} l^{\mu} m^{\nu}\,, \qquad \Phi_{10} = - \frac{1}{2} R_{\mu \nu} l^{\mu} \bar{m}^{\nu}\,,
\\
\Lambda &= \frac{1}{12} R_{\mu \nu} \left(l^{\mu} n^{\nu} - m^{\mu} \bar{m}^{\nu}\right)\,,
\\
\label{eq:phi_12}
\Phi_{12} &= -\frac{1}{2} R_{\mu \nu} n^{\mu} m^{\nu}\,, \qquad \Phi_{21} = - \frac{1}{2} R_{\mu \nu} n^{\mu} \bar{m}^{\nu}\,.
\end{align}
\end{subequations}
When the Einstein equations are imposed, this latter set of curvature scalars can be related
to the stress energy tensor $T_{\mu \nu}$ of matter through
the trace-reversed field equations:
\begin{equation}
R_{\mu \nu} = 8\pi \left(T_{\mu \nu} - \frac{1}{2} g_{\mu \nu} T\right)\,,
\end{equation}
where $T = {T_{\mu}}^{\mu}$.

%The equations governing the relationship between the spin coefficients and the Weyl scalars is found by using the relationship between the Weyl tensor and the Riemann tensor:
The decomposition of the Riemann tensor in terms of the Weyl and Ricci tensors
provide the necessary transport equations describing the evolution of the spin coefficients in terms of the above quantities; specifically
\allowdisplaybreaks[4]
\begin{widetext}
\begin{subequations}
\begin{align}
\label{eq:riem-1}
{D}\rho - \bar{{\delta}} \kappa &= (\rho^{2} + \sigma \bar{\sigma}) + \rho (\epsilon + \bar{\epsilon}) - \bar{\kappa} \tau - \kappa (3\alpha + \bar{\beta} - \pi) + \Phi_{00}\,,
\\
\label{eq:riem-2}
{D}\sigma - {\delta}\kappa &= \sigma(\rho + \bar{\rho} + 3 \epsilon - \bar{\epsilon}) - \kappa(\tau - \bar{\pi} + \bar{\alpha} + 3\beta) + \Psi_{0}
\\
\label{eq:riem-3}
{D}\tau - {\Delta}\kappa &= \rho(\tau + \bar{\pi}) + \sigma(\bar{\tau} + \pi) + \tau(\epsilon - \bar{\epsilon}) - \kappa(3\gamma + \bar{\gamma}) + \Psi_{1} + \Phi_{01}
\\
\label{eq:riem-4}
{D}\alpha - \bar{{\delta}}\epsilon &= \alpha (\rho + \bar{\epsilon} - 2 \epsilon) + \beta \bar{\sigma} - \bar{\beta} \epsilon - \kappa \lambda - \bar{\kappa} \gamma + \pi (\epsilon + \rho) + \Phi_{10}
\\
\label{eq:riem-5}
{D}\beta - {\delta}\epsilon &= \sigma(\alpha + \pi) + \beta(\bar{\rho}-\bar{\epsilon}) - \kappa(\mu + \gamma) - \epsilon(\bar{\alpha}-\bar{\pi}) + \Psi_{1}
\\
\label{eq:riem-6}
{D}\gamma - {\Delta}\epsilon &= \alpha(\tau+\bar{\pi}) + \beta(\bar{\tau}+\pi) - \gamma(\epsilon + \bar{\epsilon}) - \epsilon(\gamma + \bar{\gamma}) + \tau \pi - \nu \kappa + \Psi_{2} + \Phi_{11} - \Lambda
\\
\label{eq:riem-7}
{D}\lambda - \bar{{\delta}}\pi &= (\rho \lambda + \bar{\sigma} \mu) + \pi (\pi + \alpha - \bar{\beta}) - \nu \bar{\kappa} - \lambda (3\epsilon - \bar{\epsilon}) + \Phi_{20}
\\
\label{eq:riem-8}
{D}\mu - {\delta}\pi &= (\bar{\rho}\mu + \sigma \lambda) + \pi(\bar{\pi} - \bar{\alpha} + \beta) - \mu(\epsilon+\bar{\epsilon}) - \nu \kappa + \Psi_{2} + 2 \Lambda
\\
\label{eq:riem-9}
{D}\nu - {\Delta}\pi &= \mu(\pi + \bar{\tau}) + \lambda(\bar{\pi} + \tau) + \pi(\gamma-\bar{\gamma}) - \nu(3\epsilon+\bar{\epsilon}) + \Psi_{3} + \Phi_{21}
\\
\label{eq:riem-10}
{\Delta}\lambda - \bar{{\delta}}\nu &= -\lambda(\mu + \bar{\mu} + 3\gamma - \bar{\gamma}) + \nu(3\alpha + \bar{\beta} + \pi - \bar{\tau}) - \Psi_{4}
\\
\label{eq:riem-11}
{\delta}\rho - \bar{{\delta}}\sigma &= \rho(\bar{\alpha} + \beta) - \sigma(3\alpha-\bar{\beta}) + \tau(\rho-\bar{\rho}) + \kappa(\mu-\bar{\mu}) - \Psi_{1} + \Phi_{01}
\\
\label{eq:riem-12}
{\delta}\alpha - \bar{{\delta}}\beta &= \mu \rho - \lambda \sigma + \alpha \bar{\alpha} + \beta \bar{\beta} - 2\alpha \beta + \gamma(\rho - \bar{\rho}) + \epsilon(\mu - \bar{\mu}) - \Psi_{2} + \Phi_{11} + \Lambda
\\
\label{eq:riem-13}
{\delta}\lambda - \bar{{\delta}}\mu &= \nu(\rho-\bar{\rho}) + \pi (\mu - \bar{\mu}) + \mu (\alpha + \bar{\beta}) + \lambda(\bar{\alpha} - 3\beta) - \Psi_{3} + \Phi_{21}
\\
\label{eq:riem-14}
{\delta}\nu - {\Delta}\mu &= (\mu^{2} + \lambda \bar{\lambda}) + \mu (\gamma + \bar{\gamma}) - \bar{\nu} \pi + \nu (\tau - 3 \beta - \bar{\alpha}) + \Phi_{22}
\\
\label{eq:riem-15}
{\delta}\gamma - {\Delta}\beta &= \gamma(\tau - \bar{\alpha} - \beta) + \mu \tau - \sigma \nu - \epsilon \bar{\nu} - \beta (\gamma - \bar{\gamma} - \mu) + \alpha \bar{\lambda} + \Phi_{12}
\\
\label{eq:riem-16}
{\delta}\tau - {\Delta}\sigma &= (\mu \sigma + \bar{\lambda}\rho) + \tau (\tau + \beta - \bar{\alpha}) - \sigma (3\gamma - \bar{\gamma}) - \kappa \bar{\nu} + \Phi_{02}
\\
\label{eq:riem-17}
{\Delta}\rho - \bar{{\delta}}\tau &= -\rho \bar{\mu} + \sigma \lambda + \tau(\bar{\beta} - \alpha - \bar{\tau}) + \rho(\gamma + \bar{\gamma}) + \nu \kappa - \Psi_{2} - 2 \Lambda
\\
\label{eq:riem-18}
{\Delta}\alpha - \bar{{\delta}}\gamma &= \nu(\rho + \epsilon) - \lambda(\tau+\beta) + \alpha(\bar{\gamma} - \bar{\mu}) + \gamma(\bar{\beta} - \bar{\tau}) - \Psi_{3}
\end{align}
\end{subequations}
Meanwhile, the Bianchi identities provide the following transport equations for the Weyl scalar,
\begin{subequations}
\begin{align}
\label{eq:bianchi-1}
-\bar{{\delta}} \Psi_{0} + {D}\Psi_{1} + (4\alpha - \pi) \Psi_{0} - 2(2\rho + \epsilon) \Psi_{1} + 3 \kappa \Psi_{2} + {\cal{R}}_{a} &= 0\,,
\\
\label{eq:bianchi-2}
\bar{{\delta}}\Psi_{1} - {D} \Psi_{2} - \lambda \Psi_{0} + 2(\pi - \alpha) \Psi_{1} + 3 \rho \Psi_{2} - 2 \kappa \Psi_{3} + {\cal{R}}_{b}  &= 0\,,
\\
\label{eq:bianchi-3}
-\bar{{\delta}}\Psi_{2} + {D} \Psi_{3} + 2 \lambda \Psi_{1} - 3 \pi \Psi_{2} + 2 (\epsilon - \rho) \Psi_{3} + \kappa \Psi_{4} + {\cal{R}}_{c} &= 0\,,
\\
\label{eq:bianchi-4}
\bar{{\delta}}\Psi_{3} - {D} \Psi_{4} - 3 \lambda \Psi_{2} + 2 (2\pi + \alpha) \Psi_{3} - (4\epsilon - \rho) \Psi_{4} + {\cal{R}}_{d} &= 0\,,
\\
\label{eq:bianchi-5}
-{\Delta} \Psi_{0} + {\delta}\Psi_{1} + (4\gamma - \mu)\Psi_{0} - 2(2\tau + \beta) \Psi_{1} + 3\sigma \Psi_{2} + {\cal{R}}_{e} &= 0
\\
\label{eq:bianchi-6}
-{\Delta} \Psi_{1} + {\delta} \Psi_{2} + \nu \Psi_{0} + 2(\gamma-\mu)\Psi_{1} - 3\tau \Psi_{2} + 2\sigma \Psi_{3} + {\cal{R}}_{f} &= 0\,,
\\
\label{eq:bianchi-7}
-{\Delta} \Psi_{2} + {\delta} \Psi_{3} + 2 \nu \Psi_{1} - 3\mu \Psi_{2} + 2 (\beta - \tau) \Psi_{3} + \sigma \Psi_{4} + {\cal{R}}_{g} &= 0\,,
\\
\label{eq:bianchi-8}
-{\Delta} \Psi_{3} + {\delta} \Psi_{4} + 3 \nu \Psi_{2} - 2(\gamma + 2\mu) \Psi_{3} - (\tau - 4\beta)\Psi_{4} + {\cal{R}}_{h} &= 0\,,
\end{align}
\end{subequations}
where the ${\cal{R}}$ terms only depend on the Ricci scalars 
\begin{subequations}
\begin{align}
{\cal{R}}_{a} &= -{D}\Phi_{01} + {\delta} \Phi_{00} + 2(\epsilon + \bar{\rho})\Phi_{01} + 2 \sigma \Phi_{10} - 2 \kappa \Phi_{11} - \bar{\kappa} \Phi_{02} + (\bar{\pi} - 2 \bar{\alpha} -2 \beta) \Phi_{00}\,,
\\
{\cal{R}}_{b} &= \bar{{\delta}} \Phi_{01} - {\Delta} \Phi_{00} - 2 (\alpha + \bar{\tau}) \Phi_{01} + 2\rho \Phi_{11} + \bar{\sigma} \Phi_{02} - (\mu - 2 \gamma - 2 \bar{\gamma}) \Phi_{00} - 2\tau \Phi_{10} - 2 {D}\Lambda\,,
\\
{\cal{R}}_{c} &= -{D}\Phi_{21} + {\delta}\Phi_{20} + 2(\bar{\rho} - \epsilon) \Phi_{21} - 2\mu \Phi_{10} + 2\pi \Phi_{11} - \bar{\kappa} \Phi_{22} - (2\bar{\alpha} - 2\beta - \bar{\pi}) \Phi_{20} - 2 \bar{{\delta}} \Lambda\,,
\\
{\cal{R}}_{d} &= -{\Delta}\Phi_{20} + \bar{{\delta}} \Phi_{21} + (2\alpha - \bar{\tau}) \Phi_{21} + 2 \nu \Phi_{10} + \bar{\sigma} \Phi_{22} - 2 \lambda \Phi_{11} - (\bar{\mu} + 2\gamma - 2\bar{\gamma}) \Phi_{20}\,,
\\
{\cal{R}}_{e} &= -{D} \Phi_{02} + {\delta} \Phi_{01} + 2(\bar{\pi} - \beta) \Phi_{01} - 2\kappa \Phi_{12} - \bar{\lambda} \Phi_{00} + 2\sigma \Phi_{11} + (\bar{\rho} + 2\epsilon - 2\bar{\epsilon})\Phi_{02}\,,
\\
{\cal{R}}_{f} &= {\Delta}\Phi_{01} - \bar{{\delta}}\Phi_{02} + 2(\bar{\mu} - \gamma)\Phi_{01} - 2\rho \Phi_{12} - \bar{\nu} \Phi_{00} + 2\tau \Phi_{11} + (\bar{\tau} -2\bar{\beta} + 2\alpha)\Phi_{02} + 2{\delta} \Lambda\,,
\\
{\cal{R}}_{g} &= - {D}\Phi_{22} + {\delta}\Phi_{21} + 2(\bar{\pi} + \beta) \Phi_{21} - 2\mu \Phi_{11} - \bar{\lambda}\Phi_{20} + 2\pi \Phi_{12} + (\bar{\rho} - 2\epsilon - 2\bar{\epsilon})\Phi_{22} - 2{\Delta}\Lambda\,,
\\
{\cal{R}}_{h} &= {\Delta}\Phi_{21} - \bar{{\delta}}\Phi_{22} + 2(\bar{\mu} + \gamma) \Phi_{21} - 2\nu \Phi_{11} - \bar{\nu} \Phi_{20} + 2\lambda \Phi_{12} + (\bar{\tau} - 2\alpha - 2\bar{\beta})\Phi_{22}\,.
\end{align}
\end{subequations}
Finally, the evolution equations for the Ricci scalars are obtained through the divergence free
property of the Einstein tensor $\nabla_{\mu} G^{\mu \nu} = 0$:
\begin{subequations}
\begin{align}
\bar{{\delta}} \Phi_{01} + {\delta}\Phi_{10} - {D}(\Phi_{11} + 3\Lambda) - {\Delta}\Phi_{00} &= \bar{\kappa}\Phi_{12} + \kappa \Phi_{21} + (2\alpha +2\bar{\tau} - \pi)\Phi_{01} 
\nn \\
&+ (2\bar{\alpha} + 2\tau - \bar{\pi})\Phi_{10} - 2(\rho + \bar{\rho}) \Phi_{11} - \bar{\sigma} \Phi_{02} - \sigma \Phi_{20} 
\nn \\
&+ [\mu + \bar{\mu} - 2(\gamma + \bar{\gamma})]\Phi_{00}\,,
\\
\bar{{\delta}}\Phi_{12} + {\delta}\Phi_{21} - {\Delta}(\Phi_{11} + 3\Lambda) - {D}\Phi_{22} &= -\nu \Phi_{01} - \bar{\nu}\Phi_{10} + (\bar{\tau} - 2\bar{\beta} - 2\pi) \Phi_{12} 
\nn \\
&+ (\tau - 2\beta - 2\bar{\pi})\Phi_{21} + 2(\mu + \bar{\mu}) \Phi_{11} 
\nn \\
&- (\rho + \bar{\rho} - 2\epsilon - 2\bar{\epsilon})\Phi_{22} + \lambda \Phi_{02} + \bar{\lambda}\Phi_{20}
\\
{\delta}(\Phi_{11} - 3 \Lambda) - {D}\Phi_{12} - {\Delta}\Phi_{01} + \bar{{\delta}}\Phi_{02} &= \kappa \Phi_{22} - \bar{\nu}\Phi_{00} + (\bar{\tau} - \pi + 2 \alpha - 2\bar{\beta})\Phi_{02} - \sigma \Phi_{21} 
\nn \\
&+ \bar{\lambda}\Phi_{10} + 2(\tau - \bar{\pi})\Phi_{11} - (2\rho + \bar{\rho} - 2 \bar{\epsilon})\Phi_{12} 
\nn \\
&+ (2\bar{\mu} + \mu - 2\gamma)\Phi_{01}
\end{align}
\end{subequations}
\end{widetext}
%
%We have now enumerated all of the equations that act as the foundation of the NP formalism.

%%%%%%%%%%%%%%%%%%%%%%%%%%%%%%%%%%%%%%%%%%%%%%%%%%%%%%%%%%%%%%%%%%%%%%%%%%%%%%%%
\section{Master equations for perturbations of a Petrov Type D spacetime}
\label{section-master-equation}

   Here we review the derivation of the equations governing
the first and second order perturbations of a Petrov Type D spacetime
satisfying the vacuum Einstein equations.
The equation for first order perturbations was originally derived by
Teukolsky~\cite{Teukolsky:1973ha},
and was later generalized to $n$-th order perturbations by
Campanelli \& Lousto~\cite{Campanelli:1998jv}.
We recall that a spacetime is a Petrov Type D spacetime
if it admits two double principal null directions, with respect to which
\begin{align}
\Psi_0=\Psi_1=\Psi_3=\Psi_4=0\,.
\end{align}
By the Goldberg-Sachs theorem~\cite{GoldbergSachs},
we also have
\begin{align}
\kappa=\sigma=\nu=\lambda=0\,,
\end{align}
Finally, if the outgoing null vector $l^{\mu}$ is chosen to be
affinely parameterized, then we have, additionally, $\epsilon = 0$.
We distinguish between the background quantities and perturbations with
superscripts.
For example, for the Weyl curvature component $\Psi_0$,
we consider perturbations of the form
\begin{align}
\Psi_0= \Psi_0^{(0)}+\zeta \Psi_0^{(1)}+\zeta^{2} \Psi_0^{(2)} + {\cal{O}}(\zeta^{3})
\end{align}
where $\zeta$ is an order keeping parameter,
$\Psi_0^{(0)}$ denotes the background value,
$\Psi_0^{(1)}$ denotes the first order perturbations and $\Psi_0^{(2)}$
denotes the second order perturbation.
We similarly have second order perturbations of all the Weyl curvature,
Ricci coefficients and differential derivatives in the NP formalism.
Since the background spacetime is of Petrov Type D we have
\begin{align}
\Psi_0^{(0)}& =\Psi_1^{(0)}=\Psi_3^{(0)}=\Psi_4^{(0)}
\nn \\
&=\kappa^{(0)}=\sigma^{(0)}=\nu^{(0)}=\lambda^{(0)}=0\,.
\end{align}
By virtue of the fact that the spacetime satisfies the vacuum field equations,
the Ricci scalars in Eqs.~\eqref{eq:phi_00}-\eqref{eq:phi_12}
all vanish on the background.
For generality, we do allow these scalars to be non-zero at
first and second order in perturbation theory.

%-------------------
\subsection{First Order Perturbations}
Consider the Bianchi identities
Eqs.~\eqref{eq:bianchi-4} \& \eqref{eq:bianchi-8},
and the Riemann identity Eq.~\eqref{eq:riem-10} which can be written as
\begin{widetext}
\begin{subequations}
\begin{align}
\label{first-equation-derivation}
\left(D+4\epsilon- \rho\right) \Psi_{4}  - \left(\bar{\delta}+2\alpha + 4\pi \right)\Psi_{3} +3\lambda\Psi_{2} &=  -{\cal{R}}_{d}\,, 
\\
\label{second-equation-derivation}
- \left(\delta+ 4\beta - \tau\right) \Psi_{4} +\left(\Delta+ 2\gamma+4\mu\right) \Psi_{3} - 3\nu\Psi_{2} &=  -{\cal{R}}_{h} 
\,,
\\
\label{third-equation-derivation}
\left(\Delta + \mu + \bar{\mu} + 3\gamma - \bar{\gamma}\right)\lambda - \left(\bar{\delta} + 3\alpha + \bar{\beta} + \pi - \bar{\tau}\right)\nu &= - \Psi_{4}
\end{align}
\end{subequations}
The quantities $\{\Psi_{4}, \Psi_{3}, \lambda, \nu\}$
and the Ricci terms $\{{\cal{R}}_{d}, {\cal{R}}_{h}\}$
all vanish on the background, and thus these equations are
``of first order smallness'',
meaning that they describe the evolution of first order quantities.
Following \cite{Campanelli:1998jv}, we define the derivatives
\begin{align}
\label{eq:def_d_3}
d_3 \equiv & \bar{\delta}+3\alpha+\bar{\beta}+4\pi-\bar{\tau}, 
\\
\label{eq:def_d_4}
d_4 \equiv & \Delta+4\mu+\bar{\mu}+3\gamma-\bar{\gamma}.
\end{align}
We act on \eqref{first-equation-derivation} with $d_4^{(0)}$
and on \eqref{second-equation-derivation} with $d_3^{(0)}$
and sum the two equations to obtain
\begin{align}
\label{eq:master_eqn}
	\left[
		d_4^{(0)}\left(D+4\epsilon-\rho\right)
	-	d_3^{(0)}\left(\delta+4\beta-\tau\right)
	\right]
	\Psi_4
	&
	\nonumber \\
+	\left[
	-	d_4^{(0)}\left(\bar{\delta}+2\alpha+4\pi\right)
	+	d_3^{(0)}\left(\Delta+2\gamma+4\mu\right)
	\right]
	\Psi_3
	&
	\nonumber \\
+	3\left[
		d_4^{(0)}\lambda
	-	d_3^{(0)}\nu
	\right]
	\Psi_2
	& = - d_{4}^{(0)}{\cal{R}}_{d} - d_{3}^{(0)} {\cal{R}}_{h}
\end{align}
\end{widetext}
So far, we have not performed any perturbative expansions,
and the above equation applies at all orders in perturbation theory.

We now show how the first order term
\footnote{Observe that the zero-th order term of equation \eqref{eq:master_eqn} is trivially satisfied since $\Psi_4^{(0)}=\Psi_3^{(0)}=\lambda^{(0)}=\nu^{(0)}=0$ in a Type D spacetime.}
of the above equation corresponds to the Teukolsky equation
for Petrov Type D spacetimes.
By expanding Eq.~\eqref{eq:master_eqn} to first order we obtain
\begin{widetext}
\begin{align}
\label{eq-derivation-2}
	\left[
		d_4^{(0)}\left(D+4\epsilon-\rho\right)^{(0)}
	-	d_3^{(0)}\left(\delta+4\beta-\tau\right)^{(0)}
	\right]
	\Psi_4^{(1)}
	&
	\nonumber \\
+	\left[
	-	d_4^{(0)}\left(\bar{\delta}+2\alpha+4\pi\right)^{(0)}
	+	d_3^{(0)}\left(\Delta+2\gamma+4\mu\right)^{(0)}
	\right]
	\Psi_3^{(1)}
	&
	\nonumber \\
	+	3\left[
		d_4^{(0)}\lambda^{(1)}
	-	d_3^{(0)}\nu^{(1)}
	\right]
	\Psi_2^{(0)}
	& = - d_{4}^{(0)}{\cal{R}}_{d}^{(1)} - d_{3}^{(0)} {\cal{R}}_{h}^{(1)}
\end{align}
\end{widetext}
where we used that
$
   \Psi_4^{(0)}=\Psi_3^{(0)}=
   \lambda^{(0)}=\nu^{(0)}={\cal{R}}_{d}^{(0)}={\cal{R}}_{h}^{(0)}=0
$.
Now observe that using
\eqref{eq:comm-4}, \eqref{eq:riem-9}, \eqref{eq:riem-13}
and \eqref{eq:riem-18},
one can prove that in a vacuum Petrov Type D spacetime
\begin{align}
	\left[
	-	d_4\left(\bar{\delta}+4\pi+2\alpha\right)
	+	d_3\left(\Delta+4\mu+2\gamma\right)
		\right]
	f
	=
	0\label{eq:type_D_commutation_relation_d3_d4}
\end{align}
for any scalar $f$.
As a result of this, the second line of
Eq.~\eqref{eq-derivation-2} now vanishes.
Also, observe that using \eqref{eq:riem-10},
\eqref{eq:bianchi-7} and \eqref{eq:bianchi-3}
for a Type D background, we can derive that
\begin{align}
	\left[
		d_4^{(0)}\lambda^{(1)}
	-	d_3^{(0)}\nu^{(1)} 
	\right]
	\Psi_2^{(0)} 
	=
-	\Psi_2^{(0)}\Psi^{(1)}_4
	.
\end{align}
Putting the above together, we obtain the Teukolsky equation
\begin{align}
\label{eq:first_order_Teukolsky}
	\mathcal{T}\Psi^{(1)}_4
	=
	{\cal{R}}_{4}^{(1)}
	,
\end{align}
where $\mathcal{T}$ is the Teukolsky operator
(\cite{Teukolsky:1973ha} Eq. (2.14))
\begin{align}
\label{eq:Teukolsky_operator}
	\mathcal{T}
	&\equiv
	\left[
		d_4^{(0)}\left(D+4\epsilon-\rho\right)^{(0)} 
	-	d_3^{(0)}\left(\delta+4\beta-\tau\right)^{(0)} 
	\right]
	\nn\\
      & - 3\Psi_2^{(0)}
	,
\end{align}
and
$
   {\cal{R}}_{4}^{(1)} 
   = 
   -d_{4}^{(0)} {\cal{R}}_{d}^{(1)} - d_{3}^{(0)}{\cal{R}}_{h}^{(1)}
$.
Eq.~\eqref{eq:first_order_Teukolsky} governs the gravitational
wave perturbations
in any type D spacetime satisfying the vacuum field equations.
A solution $\Psi^{(1)}_4$ to (\ref{eq:first_order_Teukolsky})
can represent both ingoing and outgoing radiation,
though is better adapted to describing outgoing waves far from a source.
A similar procedure can be used to obtain a decoupled equation
for $\Psi_{0}^{(1)}$, which likewise can represent both
ingoing and outgoing waves, though is better adapted
to describing the former~\cite{Teukolsky:1973ha}.

%---------------------
\subsection{Second Order Perturbations}

We now turn our attention to second order perturbations of type D spacetimes. Returning to Eq.~\eqref{eq:master_eqn}, we expand to second order to obtain 
\begin{widetext}
\begin{align}
	\left[
		d_4^{(0)}\left(D+4\epsilon-\rho\right)^{(0)} 
	-	d_3^{(0)}\left(\delta+4\beta-\tau\right)^{(0)} 
	\right]
	\Psi_4^{(2)}
	&
	\nonumber \\
	+\left[
		d_4^{(0)}\left(D+4\epsilon-\rho\right)^{(1)} 
	-	d_3^{(0)}\left(\delta+4\beta-\tau\right)^{(1)} 
	\right]
	\Psi_4^{(1)}
	&
	\nonumber \\
+	\left[
	-	d_4^{(0)}\left(\bar{\delta}+2\alpha+4\pi\right)^{(1)} 
	+	d_3^{(0)}\left(\Delta+2\gamma+4\mu\right)^{(1)} 
	\right]
	\Psi_3^{(1)}
	&
	\nonumber \\
+	3\left[
		d_4^{(0)}\lambda^{(1)} 
	-	d_3^{(0)}\nu^{(1)} 
	\right]
	\Psi_2^{(1)} 
+	3\left[
		d_4^{(0)}\lambda^{(2)}
	-	d_3^{(0)}\nu^{(2)}
	\right]
	\Psi_2^{(0)}
	& =
	-d_{4}^{(0)} {\cal{R}}_{d}^{(2)} - d_{3}^{(0)} {\cal{R}}_{h}^{(2)}
	.
\end{align}
where we used Eq.~\eqref{eq:type_D_commutation_relation_d3_d4}. 
We once again make use of Eqs.~\eqref{eq:riem-10},\eqref{eq:bianchi-7},
\&~\eqref{eq:bianchi-3} to derive
\begin{align}
	\left[
		d_4^{(0)}\lambda^{(2)}
	-	d_3^{(0)}\nu^{(2)}
	\right]
	\Psi_2^{(0)}
	= &
-	\Psi^{(0)}_2\Psi^{(2)}_4
	\nonumber \\
	&
+	\Psi^{(0)}_2\left[
	-	\left(d^{(1)}_4-3\mu^{(1)}\right)\lambda^{(1)}
	+	\left(d^{(1)}_3-3\pi^{(1)}\right)\nu^{(1)}
	\right]
\end{align}
We can thus write the second order vacuum Teukolsky equation as
\begin{align}
\label{eq:second_order_Teukolsky}
	\mathcal{T}\Psi^{(2)}_4
	=
	\mathcal{S}^{(2)}_{4} + {\cal{R}}_{4}^{(2)}
	,
\end{align}
where ${\cal{R}}_{4}^{(2)} = - d_{4}^{(0)} {\cal{R}}_{d}^{(2)} - d_{3}^{(0)} {\cal{R}}_{h}^{(2)}$, and the source term $\mathcal{S}^{(2)}_{4}$ is
\begin{align}
\label{eq:second_order_Teukolsky_source}
	\mathcal{S}^{(2)}_{4} 
	\equiv &
-	\left[
		d_4^{(0)}\left(D+4\epsilon-\rho\right)^{(1)} 
	-	d_3^{(0)}\left(\delta+4\beta-\tau\right)^{(1)} 
	\right]
	\Psi_4^{(1)}
	\nonumber \\
	&
+	\left[
		d_4^{(0)}\left(\bar{\delta}+2\alpha+4\pi\right)^{(1)} 
	-	d_3^{(0)}\left(\Delta+2\gamma+4\mu\right)^{(1)} 
	\right]
	\Psi_3^{(1)}
	\nonumber \\
	&
-	3\left[
		d_4^{(0)}\lambda^{(1)}
	-	d_3^{(0)}\nu^{(1)}
	\right]
	\Psi^{(1)}_2
	\nonumber \\
	&
+	3\Psi^{(0)}_2\left[
		\left(d^{(1)}_4-3\mu^{(1)}\right)\lambda^{(1)}
	-	\left(d^{(1)}_3-3\pi^{(1)}\right)\nu^{(1)}
	\right]
	.
\end{align}
\end{widetext}
as was derived in \cite{Campanelli:1998jv} (Eq.~(9)). In particular, the source term $	\mathcal{S}^{(2)}_{4}$ only involves derivatives of the Ricci and curvature components of the background or of the first order perturbation. Further, recall that we have not yet imposed any gauge conditions on the background or the first order terms.
%%%%%%%%%%%%%%%%%%%%%%%%%%%%%%%%%%%%%%%%%%%%%%%%%%%%%%%%%%%%%%%%%%%%%%%%%%%%%%
\section{Linearized NP spin coefficients
   in terms of the linearized metric}
\label{sec:linearized_np_scalars}
   Using a choice of tetrad first described
by Chrzanowski \cite{PhysRevD.13.806}
and the commutation relations for the Newman-Penrose (NP)
derivative operators, one can rewrite the linearized NP scalars in
terms of the linearized metric components (see Sec.~\ref{lin_tetrad_our_gauge}).
Here we provide a complete listing of these relations (compare also
to Eq. (A4) of \cite{Campanelli:1998jv}):
\begin{widetext}
\begin{subequations}
\begin{align}
\label{eq:lambda-1}
\lambda^{(1)} &= \frac{1}{2} \left[-{\Delta} + 2\left(\bar{\gamma} - \gamma\right) + \mu - \bar{\mu}\right]^{(0)} h_{\om \om} - \left(\pi + \bar{\tau}\right)^{(0)} h_{n\om}\,,
\\
\label{eq:nu-1}
\nu^{(1)} &= \frac{1}{2} \left(\bar{{\delta}} + 2\alpha - \pi + 2\bar{\beta} - \bar{\tau}\right)^{(0)}h_{nn} - \left({\Delta} + 2\gamma + \bar{\mu}\right)^{(0)} h_{n\om}\,,
\\
\label{eq:sigma-1}
\sigma^{(1)} &= \frac{1}{2} \left[{D} + 2\left(\bar{\epsilon} - \epsilon\right) + \rho - \bar{\rho}\right]^{(0)} h_{mm} - \left(\tau + \bar{\pi}\right)^{(0)} h_{lm}\,,
\\
\label{eq:gamma-1}
\gamma^{(1)} &= \frac{1}{4} \left(\bar{{\delta}} + 2\bar{\beta} - 2\pi - \bar{\tau}\right)^{(0)} h_{nm} - \frac{1}{4} \left({\delta} + 2\beta + 2\bar{\pi} + 3\tau\right)^{(0)} h_{n\om} + \frac{1}{4}\left({D} + 2\bar{\epsilon} + \rho - \bar{\rho}\right)^{(0)} h_{nn} \nonumber\\
&+ \frac{1}{4} \left(\mu - \bar{\mu} - 4\gamma\right)^{(0)} h_{ln} + \frac{1}{4} \left(\mu - \bar{\mu}\right)^{(0)} h_{m\om}\,,
\\
\label{eq:kappa-1}
\kappa^{(1)} &= \left(D - 2\epsilon - \bar{\rho}\right)^{(0)} h_{lm} - \frac{1}{2} \left(\delta - 2\bar{\alpha} - 2 \beta + \bar{\pi} + \tau\right)^{(0)} h_{ll}\,.
\\
\label{eq:mu-1}
\mu^{(1)} &= \frac{1}{2} \left(\bar{{\delta}} + 2\bar{\beta} - 2\pi - \bar{\tau}\right)^{(0)} h_{nm} - \frac{1}{2} \left({\delta} + 2\beta + \tau\right)^{(0)} h_{n\om} - \frac{1}{2} \left({\Delta} - \mu + \bar{\mu}\right)^{(0)} h_{m\om} 
\nn \\
&+ \frac{1}{2} \rho^{(0)} h_{nn} - \frac{1}{2} \left(\mu + \bar{\mu}\right)^{(0)} h_{ln}\,,
\\
\label{eq:epsilon-1}
\epsilon^{(1)} &= \frac{1}{4} \left(-{\Delta} + 2\bar{\gamma} + \mu - \bar{\mu}\right)^{(0)}h_{ll} + \frac{1}{4} \left(2{D} + \rho - \bar{\rho}\right)^{(0)} h_{ln} + \frac{1}{4} \left(-{\delta} + 2\bar{\alpha} - \bar{\pi} - 2\tau\right)^{(0)} h_{l\om} 
\nn \\
&+ \frac{1}{4} \left(\bar{{\delta}} - 2 \alpha - 3\pi - 2\bar{\tau}\right)^{(0)} h_{lm} + \frac{1}{4} \left(\rho - \bar{\rho}\right)^{(0)} h_{m\om}\,,
\\
\label{eq:rho-1}
\rho^{(1)} &= \frac{1}{2} \left({D} + \rho - \bar{\rho}\right)^{(0)} h_{m\om} - \frac{1}{2} \left({\delta} + \bar{\pi} + 2\tau - 2\bar{\alpha}\right)^{(0)} h_{l\om} + \frac{1}{2} \left(\bar{{\delta}} - \pi - 2\alpha\right)^{(0)} h_{lm} 
\nn \\
&+ \frac{1}{2} \mu^{(0)} h_{ll} + \frac{1}{2} \left(\rho - \bar{\rho}\right)h_{ln}\,,
\\
\label{eq:alpha-1}
\alpha^{(1)} &= \frac{1}{4} \left(\bar{{\delta}} + 2\alpha - \pi - \bar{\tau}\right)^{(0)} h_{m\om} - \frac{1}{4} \left({\delta} - 2\bar{\alpha} + \bar{\pi} + \tau\right)^{(0)} h_{\om \om} 
\nn \\
& - \frac{1}{4} \left({\Delta} + 4 \gamma - 2\bar{\gamma} + \bar{\mu} - 2 \mu\right)^{(0)}h_{l\om} + \frac{1}{4} \left(\bar{{\delta}} - \pi - \bar{\tau}\right)^{(0)} h_{ln} + \frac{1}{4}\left({D} - 2\epsilon - \rho - 2\bar{\rho}\right)^{(0)} h_{n\om}\,,
\\
\label{eq:beta-1}
\beta^{(1)} &= \frac{1}{4} \left({D} - 4\epsilon + 2\bar{\epsilon} + 2\rho - \bar{\rho}\right)^{(0)} h_{nm} + \frac{1}{4} \left({\delta} - \bar{\pi} - \tau\right)^{(0)} h_{ln} - \frac{1}{4} \left({\delta} - 2\beta + \bar{\pi} + \tau\right)^{(0)} h_{m\om} 
\nn \\
&- \frac{1}{4} \left({\Delta} + 2\gamma + \mu + 2\bar{\mu}\right)^{(0)} h_{lm} + \frac{1}{4} \left(\bar{{\delta}} + 2\bar{\beta} - \pi - \bar{\tau}\right)^{(0)} h_{mm}\,,
\\
\label{eq:tau-1}
\tau^{(1)} &= \frac{1}{2} \left({D} + 2\bar{\epsilon} - \bar{\rho}\right)^{(0)} h_{nm} + \frac{1}{2} \left({\Delta} - 2\gamma + \mu\right)^{(0)} h_{lm} - \frac{1}{2} \left({\delta} + \bar{\pi} + \tau\right)^{(0)} h_{ln} 
\nn \\
&- \frac{1}{2} \pi^{(0)} h_{mm} - \frac{1}{2} \bar{\pi}^{(0)} h_{m\om} \,,
\\
\label{eq:pi-1}
\pi^{(1)} &= -\frac{1}{2} \left({D} + 2\epsilon - \rho\right)^{(0)} h_{n\om} - \frac{1}{2} \left({\Delta} - 2\bar{\gamma} + \bar{\mu}\right)^{(0)} h_{l\om} + \frac{1}{2} \left(\bar{{\delta}} - \pi - \bar{\tau}\right)^{(0)} h_{ln} 
\nn \\
&- \frac{1}{2} \bar{\tau}^{(0)} h_{m\om} - \frac{1}{2} \tau^{(0)} h_{\om \om}\,.
\end{align}
\end{subequations}
\end{widetext}
%
%%%%%%%%%%%%%%%%%%%%%%%%%%%%%%%%%%%%%%%%%%%%%%%%%%%%%%%%%%%%%%%%%%%%%%%%%%%%%%
\section{Alternative metric reconstruction equations}\label{sec:alt_recon}

The metric reconstruction procedure detailed in Sec.~\ref{met-recon} is not unique in the sense that one could derive alternative equations for the metric components $h_{ll}$, $h_{lm}$, and $h_{mm}$. The reason for this is that we have more equations than are necessary to solve for these components. We here provide an alternative equation for one of these components, namely $h_{ll}$. Consider the Riemann identity in Eq.~\eqref{eq:riem-8}. Linearizing this equation, we have
\begin{align}
\left(D - \bar{\rho} + \epsilon + \bar{\epsilon}\right)^{(1)} \mu^{(0)} &= \left(\delta + \bar{\pi} - \bar{\alpha} + \beta\right)^{(0)} \pi^{(1)} 
\nn \\
&+ \left(\delta + \bar{\pi} - \bar{\alpha} + \beta\right)^{(1)} \pi^{(0)} 
\nn \\
&+ \Psi_{2}^{(1)} + 2\Lambda^{(1)}
\end{align}
The left hand side of this equation contains all of the dependence on $h_{ll}$. Expanding out the left hand side, we have
\begin{widetext}
\begin{align}
\left(D - \bar{\rho} + \epsilon + \bar{\epsilon}\right)^{(1)} \mu^{(0)} &= -\frac{\mu^{(0)}}{2} \left[\Delta - \gamma - \bar{\gamma} + \bar{\mu}\right]^{(0)}h_{ll} - \frac{1}{2} h_{ll} \Delta^{(0)} \mu^{(0)}
\nn \\
&-\frac{\mu^{(0)}}{2} \left(\delta - 2\bar{\alpha} + \bar{\pi} + 2\tau\right) h_{l\om} + \frac{\mu^{(0)}}{2} \left(\bar{\delta} - 2 \alpha - \pi\right) h_{lm}\,.
\end{align}
This can be simplified by making use of the Riemann identity in Eq.~\eqref{eq:riem-14} evaluated on the background, specifically $-\Delta^{(0)}\mu^{(0)} = (\mu^{(0)})^{2} + \mu^{(0)} \left(\gamma + \bar{\gamma}\right)^{(0)}$. Applying this, we obtain a first order transport equation for $h_{ll}$, specifically
\begin{align}
\label{eq:h_ll-alt}
\left[\Delta - 2\left(\gamma + \bar{\gamma}\right) - \mu + \bar{\mu}\right]^{(0)} h_{ll} &= - \left(\delta -2 \bar{\alpha} + \bar{\pi} + 2\tau\right)h_{l\om} + \left(\bar{\delta} -2 \alpha - \pi\right) h_{lm} 
\nn \\
&- \frac{2}{\mu^{(0)}} \left[\left(\delta + \bar{\pi} - \bar{\alpha} + \beta\right)^{(0)} \pi^{(1)} + \left(\delta + \bar{\pi} - \bar{\alpha} + \beta\right)^{(1)} \pi^{(0)} + \Psi_{2}^{(1)} + 2 \Lambda^{(1)}\right]\,.
\end{align}
\end{widetext}

Why did we not make use of this equation in our case study in Sec.~\ref{case-study}? The issue with this equation is the behavior of the source term in a $1/r$ expansion. To leading order, the terms on the right hand side of Eq.~\eqref{eq:h_ll-alt} are those containing $\pi^{(1)}$ and $\Psi_{2}^{(1)}$, and which scale as $1/r^{2}$. However, these terms exactly cancel one another, and we are left with an undetermined remainder of ${\cal{O}}(1/r^{3})$. This happens to be the same order as the $h_{lm}$ and $h_{l\om}$ terms. Thus, in order to get the correct behavior of the source term in Eq.~\eqref{eq:h_ll-alt} one would have to obtain $\pi^{(1)}$ and $\Psi_{2}^{(1)}$ to higher order in $1/r$, which in turn means that we would have to start by calculating the higher order in $1/r$ corrections to $\Psi_{4}^{(1)}$. The second order transport equation in Eq.~\eqref{eq:h_ll-recon} does not have this issue. We make use of Eq.~\eqref{eq:h_ll-alt} in our numerical computations in~\cite{numerics_paper}, where this problem does not occur as we do not make any $1/r$ approximations.

This same issue arises if one tries to compute the Weyl scalar $\Psi_{0}^{(1)}$ from the expanded Riemann identity in Eq.~\eqref{eq:Psi_0-1}. The terms containing $\sigma^{(1)}$ are the leading order terms, which scale as $1/r^{3}$, and all cancel one another with a remainder of ${\cal{O}}(1/r^{4})$, which is the same order as those terms containing $\kappa^{(1)}$. By the peeling theorem, $\Psi_{0}^{(1)} = {\cal{O}}(1/r^{5})$, and thus all ${\cal{O}}(1/r^{4})$ terms in this equation must also cancel one another. Alternatively, one can solve for the remaining Weyl scalars $\Psi_{0}^{(1)}$ and $\Psi_{1}^{(1)}$ using the Bianchi identities in Eqs.~\eqref{eq:bianchi-5} \&~\eqref{eq:bianchi-6}, respectively. Expanding these equations to first order, we have
\begin{align}
\label{eq:psi_1-recon}
\left[-\Delta + 2\left(\gamma - \mu\right)\right]^{(0)} \Psi_{1}^{(1)} &+ \left(\delta - 3\tau\right)^{(0)} \Psi_{2}^{(1)} 
\nn \\
&+ \left(\delta - 3\tau\right)^{(1)} \Psi_{2}^{(0)} = - {\cal{R}}_{f}^{(1)}\,,
\\
\label{eq:psi_0-recon}
\left(-\Delta + 4\gamma - \mu\right)^{(0)} \Psi_{0}^{(1)} &+ \left[\delta - 2\left(2\tau + \beta\right)\right]\Psi_{1}^{(1)} 
\nn \\
&+ 3 \sigma^{(1)} \Psi_{2}^{(0)} = - {\cal{R}}_{e}^{(1)}\,.
\end{align}
%
%%%%%%%%%%%%%%%%%%%%%%%%%%%%%%%%%%%%%%%%%%%%%%%%%%%%%%%%%%%%%%%%%%%%%%%%%%%%%%
\bibliography{../references}

\end{document}